\documentclass[11pt,british]{article}
\PassOptionsToPackage{linecolor=RoyalBlue,backgroundcolor=RoyalBlue!25,bordercolor=RoyalBlue,textsize=scriptsize}{todonotes}
\usepackage[utf8]{inputenc}
\usepackage[dvipsnames]{xcolor}
\definecolor{lyxnotefontcolor}{rgb}{0.667969, 0, 0}
\usepackage{colortbl}
\usepackage{babel}
\usepackage{booktabs}
\usepackage{mathtools}
\usepackage{todonotes}
\usepackage{amsmath}
\usepackage{amsthm}
\usepackage{amssymb}
\usepackage{cancel}
\usepackage{geometry}
\usepackage{setspace}
\usepackage{microtype}
\usepackage[bookmarks=true,bookmarksnumbered=false,bookmarksopen=false,
 breaklinks=false,pdfborder={0 0 0},pdfborderstyle={},backref=false,colorlinks=true]
 {hyperref}
\hypersetup{pdftitle={A real superpotential for M-theory on eight-manifolds},
 pdfauthor={Anthony Ashmore and Savdeep Sethi},
 colorlinks=true,urlcolor=darkblue,anchorcolor=darkblue,citecolor=darkblue,filecolor=darkblue,linkcolor=darkblue,menucolor=darkblue,linktocpage=true}

\makeatletter

\usepackage{tikz}
\usetikzlibrary{backgrounds}
\usepackage[framemethod=tikz]{mdframed}
\AtBeginEnvironment{mdframed}{%
\tikzset{every picture/.style={}}%
}
\mdfsetup{roundcorner=.5ex,font=\small}
{\begin{mdframed}[backgroundcolor=white]}%
{\end{mdframed}}
{\begin{mdframed}[backgroundcolor=white]}%
{\end{mdframed}}

\usepackage{slashed}	 	
\usepackage{amsfonts} 		
\usepackage{cite}	
\SetTracking{encoding={*}, shape=sc}{0} 	
\allowdisplaybreaks		
\date{\today} 		
\numberwithin{equation}{section}	

\makeatletter
\g@addto@macro\bfseries{\boldmath}
\makeatother

\definecolor{darkblue}{rgb}{0.1,0.0,0.4}

\newif\ifshowrevisions
\showrevisionstrue
\newcommand{\revisioncolour}{\ifshowrevisions\color{red}\hypersetup{linkcolor=red,citecolor=red,urlcolor=red}\fi}

\makeatletter
\g@addto@macro\@floatboxreset\centering
\makeatother

\let\originalleft\left
\let\originalright\right
\renewcommand{\left}{\mathopen{}\mathclose\bgroup\originalleft}
\renewcommand{\right}{\aftergroup\egroup\originalright}

\makeatother

\begin{document}
\global\long\def\rep#1{\boldsymbol{#1}}%
\global\long\def\repb#1{\overline{\boldsymbol{#1}}}%
\global\long\def\dd{\text{d}}%
\global\long\def\ii{\text{i}}%
\global\long\def\ee{\text{e}}%
\global\long\def\Dorf{L}%
\global\long\def\tDorf{\hat{L}}%
\global\long\def\GL#1{\text{GL}(#1)}%
\global\long\def\Orth#1{\text{O}(#1)}%
\global\long\def\SO#1{\text{SO}(#1)}%
\global\long\def\Spin#1{\text{Spin}(#1)}%
\global\long\def\Symp#1{\text{Sp}(#1)}%
\global\long\def\Uni#1{\text{U}(#1)}%
\global\long\def\SU#1{\text{SU}(#1)}%
\global\long\def\Gx#1{\text{G}_{#1}}%
\global\long\def\Fx#1{\text{F}_{#1}}%
\global\long\def\Ex#1{\text{E}_{#1}}%
\global\long\def\ExR#1{\text{E}_{#1}\times\mathbb{R}^{+}}%
\global\long\def\ex#1{\mathfrak{e}_{#1}}%
\global\long\def\gl#1{\mathfrak{gl}_{#1}}%
\global\long\def\SL#1{\text{SL}(#1)}%
\global\long\def\Stab{\operatorname{Stab}}%
\global\long\def\vol{\operatorname{vol}}%
\global\long\def\tr{\operatorname{tr}}%
\global\long\def\ad{\operatorname{ad}}%
\global\long\def\ext{\Lambda}%
\global\long\def\AdS#1{\text{AdS}_{#1}}%
\global\long\def\op#1{\operatorname{#1}}%
\global\long\def\im{\operatorname{im}}%
\global\long\def\re{\operatorname{re}}%
\global\long\def\eqspace{\mathrel{\phantom{{=}}{}}}%
\global\long\def\bZ{\mathbb{Z}}%
\global\long\def\bC{\mathbb{C}}%
\global\long\def\bP{\mathbb{P}}%
\global\long\def\bR{\mathbb{R}}%
\global\long\def\feyn#1{\slashed{#1}}%
\global\long\def\id{\operatorname{id}}%
\global\long\def\ap{\alpha'}%
\global\long\def\del{\partial}%
\global\long\def\bdel{\bar{\partial}}%
\let\oldstar\star\renewcommand{\star}{\mathop{}\mathopen{}{\oldstar}}
\global\long\def\transpose{{\scriptscriptstyle \mathsf{T}}}%
\global\long\def\charge{\text{c}}%

\begin{titlepage}
\begin{flushright} 
\end{flushright}
\vfill
\begin{center}
{\setstretch{1.3}\Large\bf A real superpotential for M-theory on eight-manifolds\par} 
\vskip 1cm 
Anthony Ashmore\textsuperscript{a} and Savdeep Sethi\textsuperscript{b}
\vskip 1cm
\textit{\small{}\textsuperscript{a}Department of Physics,\\ Skidmore College, Saratoga Springs, NY 12866, USA}
\\[.5cm]
\textit{\small{}\textsuperscript{b}Leinweber Institute for Theoretical Physics \& Enrico Fermi Institute\\ \& Kadanoff Center for Theoretical Physics,\\ University of Chicago, Chicago, IL 60637, USA}
\\[.5cm]
\textit{\tt aashmore@skidmore.edu, sethi@uchicago.edu}
\end{center}
\vskip 0.1in
\begin{center} \textbf{Abstract} \end{center}
\begin{quote}
{We study warped compactifications of eleven-dimensional supergravity on eight-manifolds to Minkowski or AdS vacua preserving $N=1$ supersymmetry in three dimensions. For backgrounds characterised by a local G$_2$ structure, we derive an off-shell real superpotential functional in terms of the internal geometry and fluxes without performing a finite Kaluza--Klein truncation. We obtain this expression in two independent ways: from the dimensional reduction of the gravitino variation and from the reduction of the eleven-dimensional gravitino action. We show that the critical-point equations, together with the vacuum value of the superpotential, are equivalent to the full set of supersymmetry conditions implied by the eleven-dimensional Killing spinor equations. When the local G$_2$ structure enhances to a global Spin(7) structure, our result reduces to the known superpotential for M-theory on Spin(7) manifolds.}

\end{quote}
\vfill
\end{titlepage}

\microtypesetup{protrusion=false} 
\tableofcontents 
\microtypesetup{protrusion=true} 


\clearpage{}

\section{Introduction}

Supersymmetric flux compactifications in string theory and M-theory exhibit a rich interplay between geometry and effective field theory. On the one hand, one can work directly in ten or eleven dimensions and characterise supersymmetric solutions in terms of Killing spinors, $G$-structures and their intrinsic torsion. On the other hand, one can describe the same physics in a lower-dimensional supergravity theory, where supersymmetric vacua are controlled by a superpotential. In four-dimensional $N=1$ theories the supersymmetric couplings are encoded by a holomorphic superpotential together with a Kähler potential. In three-dimensional $N=1$ theories the scalar potential is instead determined by a \emph{real} superpotential together with a Riemannian metric on the scalar field space.

In this paper, we study M-theory, or more precisely eleven-dimensional supergravity, on an eight-manifold $X$. We focus on warped backgrounds of the form $\text{Mink}_{3}\times X$ or $\text{AdS}_{3}\times X$ preserving $N=1$ supersymmetry in three dimensions, corresponding to two real supercharges. These geometries were analysed in detail in \cite{hep-th/0511047,Martelli:2003ki}, where the necessary and sufficient conditions for supersymmetry were expressed in terms of a local $\Gx 2$-structure and its intrinsic torsion. The differential conditions were also interpreted as calibration conditions for dyonic M5-branes and used to construct explicit examples of supersymmetric solutions. The global geometry of these backgrounds, including the loci where the internal spinor becomes chiral, has also been studied using foliations and singular foliations~\cite{1411.3148,1411.3497}.

Our main result is the construction of the real superpotential functional governing the resulting three-dimensional $N=1$ theory. This provides the analogue, for this class of M-theory backgrounds, of the Gukov--Vafa--Witten superpotential familiar from other flux compactifications~\cite{Gukov:1999ya}. We show that supersymmetric solutions are extrema of this functional. Conversely, by imposing the vanishing of the first variation together with the appropriate vacuum value of the superpotential, we recover the full set of supersymmetry conditions on the internal geometry implied by the eleven-dimensional Killing spinor equations. In this sense, the higher-dimensional supersymmetry conditions are the critical-point equations of a three-dimensional superpotential. 

We derive the superpotential in two independent ways. The first reads it off from the dimensional reduction of the gravitino variation, following for example \cite{hep-th/0507104,Becker:2002jj,PiresPacheco:2008qik}. The second computes the three-dimensional gravitino mass by reducing the eleven-dimensional gravitino action, in the spirit of \cite{hep-th/0408121,House:2004pm,deIaOssa:2019cci}. The agreement between these two derivations provides a non-trivial consistency check. We also show that in the limit where the local $\Gx 2$ structure enhances to a global $\Spin 7$ structure, our expression reduces to the known superpotential for M-theory on $\Spin 7$ manifolds~\cite{Gukov:1999gr,Acharya:2002vs,Becker:2002jj}, while the supersymmetry conditions are recovered only by varying the general superpotential before restricting to the $\Spin 7$ locus.

A central feature of our approach is that the superpotential we derive is off-shell. For a general flux compactification, it is not known how to determine the full Kaluza--Klein spectrum, and so usually one cannot integrate out the massive modes to obtain a conventional effective theory with finitely many fields. Instead, using the perspective of \cite{deWit:1986mz,hep-th/0602241,Grana:2005ny,Grana:2006hr}, we formally retain the full tower of Kaluza--Klein modes while singling out two real supercharges. This motivates an off-shell superpotential functional on the space of eleven-dimensional geometric and flux data, whose critical-point equations recover the higher-dimensional supersymmetry conditions. Here ``off-shell'' means that the functional is defined without imposing the Killing spinor equations; we do not construct a complete off-shell formulation of the infinite Kaluza--Klein theory. Once a suitable light sector is identified, this functional provides the starting point for deriving its effective superpotential.

This perspective is especially useful in flux compactifications, where the relation between the higher-dimensional geometry and a truncated effective theory is often subtle. In many familiar treatments, one begins with a fluxless compactification, identifies the massless modes there, and then incorporates fluxes by adding flux-dependent terms to the lower-dimensional superpotential. While this procedure is often powerful, it may miss information about the full higher-dimensional geometry -- see, for example, \cite{hep-th/0509003} or \cite[Section 5]{0707.1038}. Moreover, since backgrounds with and without fluxes often have drastically different geometric and even topological properties, it is often unclear whether the effective theory derived from the fluxless compactification is trustworthy after turning on fluxes~\cite{Sethi:2017phn}. Our analysis avoids this issue by working directly with the geometry of the flux background itself and deriving the superpotential before any truncation to a finite set of modes.

The present work fits into the broader programme of relating the direct analysis of supersymmetric solutions in ten or eleven dimensions to their description in lower-dimensional effective supergravity. Similar superpotential calculations have been carried out in a wide variety of settings, such as type II compactifications on Calabi--Yau threefolds~\cite{Gukov:1999ya,Gukov:1999gr,hep-th/9912152}, $\SU 3\times\SU 3$-structure manifolds~\cite{Grana:2005ny,Grana:2006hr,0707.1038}, M-theory on Calabi--Yau fourfolds~\cite{Gukov:1999gr,hep-th/0103068}, $\Spin 7$ manifolds~\cite{Gukov:1999gr,Acharya:2002vs}, including the flux-dependent three-dimensional effective action of~\cite{1307.5858}, $\SU 3$ structure manifolds~\cite{hep-th/0311119}, $\Gx 2$-structure manifolds~\cite{House:2004pm,PiresPacheco:2008qik}, heterotic compactifications on half-flat manifolds~\cite{hep-th/0408121} and $\Gx 2$ structure manifolds~\cite{deIaOssa:2019cci}, and heterotic M-theory on $\SU 3$-structure manifolds~\cite{hep-th/0602039}. More recently, compactifications to three dimensions on $\Gx 2$-structure manifolds have been studied in type II theories at the level of both explicit vacua and finite-dimensional effective theories; see for example~\cite{2005.05246,2103.03282}. Warped throats in the type IIA setting have also been analysed in~\cite{2505.21104}. A unified treatment of three-dimensional $N=1$ effective theories obtained from ten-dimensional string theories on seven-manifolds with $\Gx 2$ structure was recently given in~\cite{2605.06800}. That analysis assumes a finite-dimensional truncation of the internal forms. A related approach has also been developed for type IIA on $\Spin7$ spaces, yielding two-dimensional $N=(1,1)$ supergravity~\cite{2608.18284}. The present work gives the off-shell superpotential for the class of warped M-theory compactifications on eight-manifolds with local $\Gx 2$ structure studied in~\cite{Martelli:2003ki,hep-th/0511047}. In particular, we derive the superpotential before selecting a finite set of modes.

The paper is organised as follows. In Section \ref{sec:Background}, we review the supersymmetry conditions for warped $\text{Mink}_{3}$ and $\text{AdS}_{3}$ backgrounds of M-theory on eight-manifolds and the associated local $\Gx 2$ structure, and collect some results about three-dimensional $N=1$ supergravity. In Section \ref{sec:Computing-the-off-shell}, we derive the off-shell superpotential from the gravitino variation and from the reduction of the eleven-dimensional gravitino action, and show that it reduces to the known superpotential for M-theory on $\Spin 7$ manifolds. In Section \ref{sec:Supersymmetry-conditions-from}, we show in detail that critical points of the superpotential reproduce the necessary and sufficient supersymmetry conditions on the internal geometry. In Section \ref{sec:The-global-}, we explain how the superpotential behaves in the $\Spin 7$ limit and clarify how the familiar $\Spin 7$ supersymmetry conditions are recovered. We conclude in Section \ref{sec:Discussion} with some interpretation of our results and some possible directions for future work. The appendices contain our conventions and further details on the Killing spinor bilinears and $\Gx 2$ structure used in the main text.

\section{Background\label{sec:Background}}

We begin by recalling the structure of minimally supersymmetric solutions to M-theory on eight-manifolds and summarise how supersymmetry can be phrased in terms of differential conditions on a local $\Gx 2$ structure, as first laid out in \cite{Martelli:2003ki}. We then outline the structure of the effective minimal supergravity in three dimensions that should result from compactification. 

\subsection{Eleven-dimensional supergravity on eight-manifolds\label{subsec:Eleven-dimensional-supergravity-}}

We follow the notation of \cite{Martelli:2003ki}. As usual, the field content of eleven-dimensional supergravity is a metric $\hat{g}$, a local three-form $C$ with field strength $G=\dd C$, and a gravitino $\Psi_{M}$. A supersymmetric background is a bosonic solution, with vanishing gravitino, that admits at least one nowhere-vanishing Majorana spinor $\eta$ for which the gravitino variation vanishes:
\begin{equation}
\delta\Psi_{M}\equiv\hat{\nabla}_{M}\eta-\tfrac{1}{288}\left(\hat{\Gamma}^{NPQR}{}_{M}G_{NPQR}-8G_{MNPQ}\hat{\Gamma}^{NPQ}\right)\eta=0,\label{eq:gravitino_variation}
\end{equation}
where the $\hat{\Gamma}_{M}$ furnish a representation of the eleven-dimensional Clifford algebra, $\{\hat{\Gamma}_{M},\hat{\Gamma}_{N}\}=2\hat{g}_{MN}$.

We require Poincar\'e or AdS invariance in the three external dimensions, and therefore take the background to be a warped product of three-dimensional Minkowski or AdS space with an eight-manifold $X$. The metric then splits as
\begin{equation}
\dd\hat{s}^{2}_{11}=\ee^{2\Delta}\left(h_{\mu\nu}\dd y^{\mu}\dd y^{\nu}+g_{mn}\dd x^{m}\dd x^{n}\right),\label{eq:11_metric}
\end{equation}
where, on-shell, $h_{\mu\nu}$ is a maximally symmetric metric on Minkowski or AdS, and $\Delta$ is a warp factor that depends only on the internal coordinates. We take $X$ to be compact and without boundary, with smooth fields. The local differential analysis that follows also applies on patches. The general form of the field strength $G$ that is compatible with this ansatz is
\begin{equation}
G=\ee^{3\Delta}(F+\vol_{3}\wedge f),\label{eq:flux_warp}
\end{equation}
where $\vol_{3}$ is the volume form associated to $h$, $F$ is a four-form on $X$, and $f$ is a one-form on $X$, so that $F$ and $f$ are the magnetic and electric components of the flux respectively.

In general, the eleven-dimensional supersymmetry parameter can be decomposed as\footnote{There appears to be a typographical error in \cite{Martelli:2003ki}, where this is written with the factor $\ee^{-\Delta/2}$. One needs to take $\ee^{\Delta/2}$ in order to reproduce the Killing spinor equations \eqref{eq:KSE-1} and \eqref{eq:KSE-2}, which are Equations (2.25) and (2.26) in \cite{Martelli:2003ki}.}
\begin{equation}
\eta=\ee^{\Delta/2}\psi\otimes(\xi_{+}+\xi_{-}),\label{eq:spinor_ansatz}
\end{equation}
where $\psi$ is a Majorana spinor in three dimensions, and $\xi_{\pm}$ are Majorana--Weyl spinors in eight dimensions. Note that, given our conventions in Appendix \ref{app:conventions}, Majorana spinors in eight dimensions are simply real spinors. For an $N=1$ background with a three-dimensional Minkowski or AdS factor, we require that the external spinor $\psi$ satisfies
\begin{equation}
\nabla_{\mu}\psi+m\gamma_{\mu}\psi=0,\label{eq:ads_killing}
\end{equation}
where $m$ is proportional to the inverse AdS radius and normalised so that the Ricci scalar for $h_{\mu\nu}$ is $R=-24m^{2}$. As noted in \cite{Martelli:2003ki}, it is convenient to define the nowhere-vanishing \emph{non-chiral} Majorana spinors
\begin{equation}\label{eq:G2-spinor-relation}
\epsilon_{+}=\frac{1}{\sqrt{2}}(\xi_{+}+\xi_{-}),\qquad\epsilon_{-}\equiv\gamma_{9}\epsilon_{+}=\frac{1}{\sqrt{2}}(\xi_{+}-\xi_{-}),
\end{equation}
where the highest-rank gamma matrix $\gamma_{9}=\gamma_{1}\dots\gamma_{8}$ acts as $\gamma_{9}\xi_{\pm}=\pm\xi_{\pm}$. 

A background is supersymmetric if it admits a nowhere-vanishing spinor $\eta$ for which the gravitino variation vanishes. Substituting the spinor ansatz \eqref{eq:spinor_ansatz} into the gravitino variation \eqref{eq:gravitino_variation} gives internal and external equations, which can be expressed in terms of $\epsilon_{\pm}$ as
\begin{align}
\nabla_{m}\epsilon_{\pm}\pm\tfrac{1}{24}F_{mpqr}\gamma^{pqr}\epsilon_{\pm}-\tfrac{1}{4}f_{n}\gamma^{n}{}_{m}\epsilon_{\mp}\pm m\gamma_{m}\epsilon_{\mp} & =0,\label{eq:KSE-1}\\
\tfrac{1}{2}\partial_{m}\Delta\gamma^{m}\epsilon_{\pm}\mp\tfrac{1}{288}F_{mpqr}\gamma^{mpqr}\epsilon_{\pm}-\tfrac{1}{6}f_{m}\gamma^{m}\epsilon_{\mp}\mp m\epsilon_{\mp} & =0.\label{eq:KSE-2}
\end{align}
We refer to the $\epsilon_{\pm}$ as Killing spinors and to these equations as the Killing spinor equations. A supersymmetric solution must also satisfy the flux equations and the Einstein equation. For the ansatz above, the Bianchi identity $\dd G=0$ has two components
\begin{equation}
\dd(\ee^{3\Delta}F)=0,\qquad \dd(\ee^{3\Delta}f)=0.
\end{equation}
As explained in \cite{Martelli:2003ki}, the second condition actually follows from the Killing spinor equations with the remaining flux conditions given by
\begin{equation}
\dd(\ee^{3\Delta}F)=0,\qquad\ee^{-6\Delta}\dd(\ee^{6\Delta}\star f)=-\tfrac{1}{2}F\wedge F,\label{eq:bianchi1}
\end{equation}
where $\star$ is the Hodge star associated to the unwarped metric $g$. The first is the magnetic Bianchi identity; the second is the purely internal component of the $G$ equation of motion, or equivalently the Bianchi identity for the dual seven-form field strength. Together with the Killing spinor equations, they imply the remaining bosonic field equations.

Our analysis uses the two-derivative, source-free supergravity equations. On a smooth compact $X$ without boundary, integrating \eqref{eq:bianchi1} gives a necessary global constraint
\begin{equation}
\frac12\int_X\mathcal F\wedge\mathcal F=0,
\qquad\mathcal F\equiv\ee^{3\Delta}F,\label{eq:compact_flux_constraint}
\end{equation}
where $\mathcal F$ is the closed internal magnetic flux. In the positive-chirality global $\Spin7$ supersymmetric limit considered in Section~\ref{sec:The-global-}, this flux is self-dual, so the left-hand side has positive norm and \eqref{eq:compact_flux_constraint} forces $\mathcal F=0$. Thus, the local differential conditions are not sufficient for the existence of compact flux vacua. Sources or higher-derivative corrections are usually invoked to avoid this no-go theorem. For example, with the usual $X_8$ term defined in Appendix~\ref{subsec:Eleven-dimensional-supergravity} and no space-filling M2-branes, equation \eqref{eq:bianchi1} is corrected to
\begin{equation}
\frac12\int_X\mathcal F\wedge\mathcal F = -\beta\int_X X_8,
\label{eq:compact_X8_constraint}
\end{equation}
which now allows non-trivial flux. Rather than picking and choosing ingredients, flux quantisation, branes and higher-derivative corrections should be treated together when studying the existence of a compact M-theory background with flux~\cite{Becker:1996gj, Sethi:1996es, Sethi:2017phn}. We will defer an analysis of this global constraint for future work.

\subsection{Supersymmetry and a local \texorpdfstring{$\Gx 2$}{G\_2} structure\label{subsec:Martelli_review}}

The nowhere-vanishing Majorana spinor $\epsilon_{+}$ defines a local $\Gx 2$ structure in eight dimensions. The decomposition of the space of $p$-forms on $X$ under $\Gx 2\subset\SO 8$ shows that such a structure defines an invariant one-form, three-form and four-form, together with the Hodge duals. In addition, the relative orientation of $\epsilon_{+}$ and $\epsilon_{-}$ defines a function on $X$. 

Given the choice of warp factor in \eqref{eq:spinor_ansatz}, the Killing spinor equations imply that the norms of $\epsilon_{+}$ and $\epsilon_{-}$ are constant. 
To see this, contract the first Killing spinor equation \eqref{eq:KSE-1} with $\epsilon_+^\transpose$ and add the transposed equation. The contribution from the four-form flux is
\begin{equation}
	-\tfrac{1}{24}F_{mpqr}\epsilon_+^\transpose\gamma^{pqr}\epsilon_+
	+\tfrac{1}{24}F_{mpqr}\epsilon_+^\transpose\gamma^{pqr}\epsilon_+ = 0 ,
\end{equation}
where the relative sign follows from the antisymmetry of $\gamma_{(3)}$ -- no condition on the three-form bilinear is needed. Thus
\begin{equation}
	\nabla_m(\epsilon_+^\transpose\epsilon_+)
	=
	\tfrac{1}{2} f^n \epsilon_+^\transpose\gamma_{nm}\epsilon_-
	-2m\,\epsilon_+^\transpose\gamma_m\epsilon_- .
\end{equation}
Using $\epsilon_-=\gamma_9\epsilon_+$, the remaining terms are proportional to $\epsilon_+^\transpose\gamma_{nm}\gamma_9\epsilon_+$ and $\epsilon_+^\transpose\gamma_m\gamma_9\epsilon_+$. These vanish by the identities in \eqref{eq:majorana_identities}, or equivalently by the rank-two and rank-one
mixed-bilinear entries of Table~\ref{tab:Rank--gamma-matrix}. Hence	$\nabla_m(\epsilon_+^\transpose\epsilon_+)=0$. The same argument with $+\leftrightarrow-$ gives $\nabla_m(\epsilon_-^\transpose\epsilon_-)=0$.

We are then free to normalise the spinors as
\begin{equation}
\Vert\epsilon_{\pm}\Vert^{2}=\epsilon^{\transpose}_{\pm}\epsilon_{\pm}=1.
\end{equation}
In general, $\epsilon^{\transpose}_{+}\epsilon_{-}$ is not constant and takes values in $[-1,1]$. This is parametrised by an angle $\zeta$ as
\begin{equation}
\epsilon^{\transpose}_{+}\epsilon_{-}\equiv\sin\zeta,
\end{equation}
which can vary over $X$. The invariant one-form, three-form and four-form are given by bilinears of $\epsilon_{\pm}$ as
\begin{equation}
\bar{K}=\epsilon^{\transpose}_{+}\gamma_{(1)}\epsilon_{+},\qquad\bar{\phi}=\epsilon^{\transpose}_{-}\gamma_{(3)}\epsilon_{+},\qquad Y=\epsilon^{\transpose}_{+}\gamma_{(4)}\epsilon_{+}.\label{eq:invariant_forms}
\end{equation}
Using Fierz identities, one finds that $\imath_{\bar{K}}\bar{\phi}=0$. 

Note that the $\Gx 2$ structure is only \emph{locally} defined since there can be points on $X$ where the Killing spinors are stabilised by a group other than $\Gx 2$. Specifically, at points where $\sin\zeta=\pm1$, $\epsilon_{+}$ and $\epsilon_{-}$ are aligned or anti-aligned (positive or negative chirality), so both $\bar{K}$ and $\bar{\phi}$ vanish. The four-form $Y$ is still non-zero and defines a $\Spin 7$ structure. With a fixed orientation on $X$, $Y$ satisfies $\star Y=sY$ on the branch $\sin\zeta=s=\pm1$; it is self-dual on the positive-chirality branch.\footnote{If $\sin\zeta=\pm1$ everywhere so that the $\Spin 7$ structure is global, the analysis reduces to that given in \cite{Acharya:2002vs}. If the spinor is everywhere nonchiral, the closed one-form determines a codimension-one foliation carrying a leafwise $\Gx2$ structure~\cite{1411.3148}. Allowing a chiral locus leads to a singular-foliation description~\cite{1411.3497}.} 

Away from these loci, one can introduce a pair of rescaled forms
\begin{equation}
K=\frac{1}{\cos\zeta}\bar{K},\qquad\phi=\frac{1}{\cos\zeta}\bar{\phi},
\end{equation}
which are normalised as $\Vert K\Vert^{2}=1$ and $\Vert\phi\Vert^{2}=7$, and so define a canonical $\Gx 2$ structure on $X$. The four-form $Y$ is not independent and can be written as
\begin{equation}
Y=-\psi+\sin\zeta\,\phi\wedge K,
\end{equation}
where $\psi=\star_{7}\phi$, the seven-dimensional Hodge star is defined as $\star_{7}=\imath_{K}\star$, and $\star$ is the Hodge star associated to $g$. These forms may be written explicitly in a convenient orthonormal frame, as in \cite{Gauntlett:2003cy}. We give details of this in Appendix \ref{app:G2-bilinears}.

The Killing spinor equations impose differential conditions on the spinor bilinears. Combining \eqref{eq:KSE-1} and \eqref{eq:KSE-2} with some standard gamma-matrix algebra, one can derive a set of derivatives and identities, which we give in Appendix \ref{app:bilinears}. A necessary and sufficient set of conditions for a solution to the Killing spinor equations is~\cite{Martelli:2003ki}
\begin{align}
\dd(\ee^{3\Delta}\cos\zeta\,K) & =0,\label{eq:sufficient_1}\\
K\wedge\dd(\ee^{6\Delta}\psi) & =0,\\
\ee^{-12\Delta}\dd(\ee^{12\Delta}\cos\zeta\vol_{7}) & =-8m\sin\zeta\vol_{7}\wedge K,\\
\cos\zeta\,\dd\phi\wedge\phi & =32m\vol_{7}-4\star\dd\zeta+2\cos\zeta\star f,\label{eq:corrected}\\
\ee^{-3\Delta}\dd(\ee^{3\Delta}\sin\zeta) & =f-4m\cos\zeta\,K,\label{eq:sufficient_5}\\
\ee^{-6\Delta}\dd(\ee^{6\Delta}\cos\zeta\,\phi) & =-\star F+\sin\zeta\,F-4mY,\label{eq:sufficient_6}
\end{align}
where we have defined $\vol_{7}\equiv\star_{7}1=\tfrac{1}{7}\phi\wedge\psi$ and $\vol_{8}\equiv\star1=K\wedge\vol_{7}$.\footnote{Equation \eqref{eq:corrected} corrects a typo in \cite{Martelli:2003ki}. It agrees with the equation derived in \cite{Legramandi:2020txf} up to a difference in convention for the Hodge star.} Note that these conditions completely determine the intrinsic torsion of the $\Gx 2$ structure, and fix the electric flux $f$ and the $\rep 1$ and $\rep 7$ $\Gx 2$ components of the magnetic flux $F$ in terms of $\dd\zeta$, $\dd\Delta$ and $m$. The two $\rep{27}$ components of $F$ are not constrained by the supersymmetry conditions, though they must still satisfy \eqref{eq:bianchi1} for a solution to the equations of motion.

\subsection{Three-dimensional \texorpdfstring{$N=1$}{N = 1} supergravity\label{subsec:Three-dimensional--supergravity}}

Compactification of eleven-dimensional supergravity on a compact eight-manifold $X$ admitting a single nowhere-vanishing Majorana spinor $\epsilon_{+}$ leads, at least formally, to a three-dimensional theory with $N=1$ supersymmetry. If $\epsilon_{+}$ also solves the Killing spinor equations, the effective theory will admit a Minkowski or AdS vacuum preserving $N=1$ supersymmetry. To see how the geometry of $X$ determines the three-dimensional effective theory, we first recall a few facts about minimally supersymmetric matter-coupled supergravity in three dimensions. Our discussion follows \cite{0907.4658,2005.05246} -- more details can be found in \cite{deWit:2003ja,Becker:2003wb,1101.4013,1710.00554}.

The bosonic sector of matter-coupled minimal supergravity in three dimensions has a metric $g_{\mu\nu}$, real scalar fields $\phi^{I}$ and abelian gauge fields $A^{(A)}$. The bosonic part of the Lagrangian is given by
\begin{equation}
\mathcal{L}=\tfrac{1}{2}R-g^{\mu\nu}G_{IJ}(\phi)\partial_{\mu}\phi^{I}\partial_{\nu}\phi^{J}-\tfrac{1}{4}f(\phi)F^{(A)}_{\mu\nu}F^{\mu\nu(A)}-V(\phi),\label{eq:3d_Action}
\end{equation}
where the $F^{(A)}$ are field strengths for the gauge fields, and the gauge kinetic functions $f(\phi)$ are real but otherwise unrestricted. The potential for the scalar fields is given by
\begin{equation}
V(\phi)=G^{IJ}\mathcal{W}_{I}\mathcal{W}_{J}-4\mathcal{W}^{2},
\end{equation}
where $\mathcal{W}(\phi)$ is a real function of the $\phi^{I}$, $\mathcal{W}_{I}\equiv\partial\mathcal{W}/\partial\phi^{I}$, and $G_{IJ}$ is the target space metric on the space of scalar fields. Thus $\mathcal{W}$ plays the role of a \emph{real} superpotential, determining the scalar potential in the same way that a superpotential and K\"ahler potential do in theories with more supersymmetry. However, unlike theories with more supersymmetry, the superpotential is not holomorphic and so is not expected to be protected in any way. Furthermore, the target space metric is generically only Riemannian and not Kähler, as would be the case for an $N=2$ theory.

The fermionic sector contains a gravitino $\psi_{\mu}$, Majorana spinors $\chi^{I}$ and Majorana gaugini $\lambda^{(A)}$. The quadratic terms in the gravitino action are
\begin{equation}
\mathcal{L}_{3/2}=-\tfrac{1}{2}\bar{\psi}_{\mu}\gamma^{\mu\nu\rho}\nabla_{\nu}\psi_{\rho}-\tfrac{1}{2}\mathcal{W}\,\bar{\psi}_{\mu}\gamma^{\mu\nu}\psi_{\nu},\label{eq:3d_fermion_action}
\end{equation}
with the fermion shifts on Minkowski or AdS given by
\begin{equation}
\delta\psi_{\mu}|_{\text{shift}}=\tfrac{1}{4}\omega_{\mu}{}^{ab}\gamma_{ab}\psi-\mathcal{W}\,\gamma_{\mu}\psi,\qquad\delta\chi^{I}|_{\text{shift}}=G^{IJ}\mathcal{W}_{J}\psi,\qquad\delta\lambda^{(A)}|_{\text{shift}}=0.\label{eq:3d_shift}
\end{equation}
Note that we have taken the local supersymmetry parameter $\psi$ to be a two-component Majorana spinor in three dimensions, in agreement with \eqref{eq:spinor_ansatz}. A supersymmetric AdS$_{3}$ vacuum satisfies
\begin{equation}
\langle\mathcal{W}_{I}\rangle=0,\qquad m_{3/2}=\langle\mathcal{W}\rangle=\frac{1}{2L_{\text{AdS}}},\qquad\langle V\rangle=-4\langle\mathcal{W}^{2}\rangle=-\frac{1}{L^{2}_{\text{AdS}}}.\label{eq:3d_susy}
\end{equation}
Following \cite{Freedman:2012zz,2005.05246}, the Killing spinor equation on AdS is defined as $(\nabla_{\mu}-\tfrac{1}{2}L^{-1}_{\text{AdS}}\gamma_{\mu})\psi=0$. Compared with \eqref{eq:ads_killing}, we see that $m^{-1}=-2L_{\text{AdS}}$. For a Minkowski vacuum, one has $L_{\text{AdS}}\to\infty$ and so also requires $m_{3/2}=\langle V\rangle=0$.

\section{Computing the off-shell \texorpdfstring{$N=1$}{N = 1} superpotential\label{sec:Computing-the-off-shell}}

Comparing the three- and eleven-dimensional perspectives, the real superpotential $\mathcal{W}$ governing supersymmetry in the three-dimensional effective theory should be determined by the geometry of and the fluxes on the eight-manifold $X$.

Since the moduli space of general $N=1$ compactifications on eight-manifolds is not well understood, it is not possible in general to identify the massless or light modes and then integrate out the massive ones so as to obtain a finite-dimensional effective supergravity in three dimensions. Instead, following \cite{Grana:2005ny,hep-th/0602241,Grana:2006hr}, we keep the full tower of Kaluza--Klein modes on $X$. The resulting superpotential should therefore encode the full supersymmetry data of the eleven-dimensional background. In particular, without truncating to the light modes, the conditions for a supersymmetric vacuum in three dimensions, given in \eqref{eq:3d_susy}, should reproduce the eleven-dimensional supersymmetry conditions encoded in the Killing spinor equations \eqref{eq:KSE-1} and \eqref{eq:KSE-2}. This perspective has been used for backgrounds in type II and heterotic supergravity in, for example, \cite{0707.1038, Grana:2005ny, Grana:2006hr, PiresPacheco:2008qik, Grana:2009im, delaOssa:2015maa, McOrist:2016cfl, Ashmore:2019rkx, Ashmore:2019qii, deIaOssa:2019cci}.

We emphasise that our objective is to determine the off-shell superpotential itself, not merely its value on a supersymmetric vacuum. For this reason, it is not enough to infer $\mathcal{W}$ only from on-shell data such as the tension of a BPS domain wall from a brane wrapping an internal cycle, as is often done in type II compactifications~\cite{Gukov:1999ya,Haack:2009jg}. 

There are two natural ways to determine the off-shell superpotential: dimensionally reduce the eleven-dimensional fermion action and read off the gravitino mass in three dimensions, or reduce the gravitino variation and read off the form of $\mathcal{W}$ directly. Both approaches require that the three-dimensional metric be brought to Einstein frame and that the gravitino kinetic term be canonically normalised. We begin with these preliminary steps before turning to the superpotential itself.

We start by decomposing the eleven-dimensional gravitino as
\begin{equation}
\Psi_{\mu}=\ee^{\Delta/2}\psi_{\mu}\otimes(\xi_{+}+\xi_{-})=\sqrt{2}\ee^{\Delta/2}\psi_{\mu}\otimes\epsilon_{+},
\end{equation}
which mirrors the decomposition of the eleven-dimensional spinor $\eta$ in \eqref{eq:spinor_ansatz}. As we discuss in Section \ref{sec:The--superpotential}, $\psi_{\mu}$ is not the actual three-dimensional gravitino. Instead, one needs to shift $\psi_{\mu}$ in order to diagonalise the kinetic term of the fermion Lagrangian. This will be important for reading off $\mathcal{W}$ from the gravitino variation. However, this subtlety does not affect the overall normalisation of the kinetic term or the gravitino mass term, since both are quadratic in the gravitino.

Starting from the Einstein--Hilbert term in eleven dimensions, our ansatz \eqref{eq:11_metric} for the metric and the corresponding decomposition of the eleven-dimensional gamma matrices \eqref{subsec:Metric-and-gamma} give
\begin{equation}
\frac{1}{2\kappa^{2}}\int\dd^{11}X\sqrt{-\hat{g}}\hat{R}=\frac{1}{2\kappa^{2}}\left(\int\dd^{3}y\sqrt{-h}R_{3}\right)\mathcal{V}+\dots
\end{equation}
Here $y^{\mu}$ are coordinates on the three-dimensional external space, and we have assumed that the warp factor depends only on the internal coordinates. The quantity $\mathcal{V}$ denotes the internal warped volume,
\begin{equation}
\mathcal{V}=\int\dd^{8}x\sqrt{g}\,\ee^{9\Delta}\equiv\int\ee^{9\Delta}\vol_{8}.\label{eq:warped_volume}
\end{equation}
Performing the Weyl transformation $h_{\mu\nu}\to\mathcal{V}^{-2}h_{\mu\nu}$ puts the resulting theory in three-dimensional Einstein frame.

The kinetic term for the gravitino comes from the part of the eleven-dimensional kinetic term with all indices external. One finds
\begin{equation}
-\frac{1}{\kappa^{2}}\int\dd^{11}X\sqrt{-\hat{g}}\left(\bar{\Psi}_{M}\hat{\Gamma}^{MNP}\hat{\nabla}_{N}\Psi_{P}\right)=-\frac{2}{\kappa^{2}}\left(\int\dd^{3}y\sqrt{-h}\,\bar{\psi}_{\mu}\gamma^{\mu\nu\rho}\nabla_{\nu}\psi_{\rho}\right)\mathcal{V}+\dots
\end{equation}
where we have used the ansatz \eqref{eq:gamma_decomposition} for the gamma matrices, together with $\epsilon^{\transpose}_{\pm}\epsilon_{\pm}=1$ and $\epsilon^{\transpose}_{-}\gamma^{n}\epsilon_{+}=0$. We find a canonical kinetic term that agrees with the three-dimensional fermion action \eqref{eq:3d_fermion_action} after setting $\kappa=1$ and transforming 
\begin{equation}
h_{\mu\nu}\to\mathcal{V}^{-2}h_{\mu\nu},\qquad\gamma_{\mu}\to\mathcal{V}^{-1}\gamma_{\mu},\qquad\psi_{\mu}\to\tfrac{1}{2}\mathcal{V}^{-1/2}\psi_{\mu}.\label{eq:rescalings}
\end{equation}
Compared with the Killing spinor equation \eqref{eq:ads_killing} on AdS$_{3}$ and our ansatz for the metric \eqref{eq:11_metric}, these transformations imply $m\to\mathcal{V}^{-1}m$, so that the inverse AdS scale in Einstein frame is set by $m/\mathcal{V}$. Compared with our conventions for three-dimensional supergravity in \eqref{eq:3d_susy}, the AdS radius in three-dimensional Einstein frame is given by $L_{\text{AdS}}=-\mathcal{V}/2m$. This implies that the on-shell value of the superpotential after transforming to Einstein frame is
\begin{equation}
\langle\mathcal{W}\rangle=-\frac{m}{\mathcal{V}}.\label{eq:rescaled}
\end{equation}

\subsection{Dimensional reduction of the gravitino variation\label{sec:The--superpotential}}

In this subsection, we use the supersymmetry variation of the eleven-dimensional gravitino to determine the real superpotential $\mathcal{W}$ of the resulting three-dimensional supergravity in terms of the local $\Gx 2$ structure on $X$. Our analysis follows the spirit of \cite{Becker:2002jj,hep-th/0507104,PiresPacheco:2008qik}, where one derives an expression for $\mathcal{W}$ in terms of invariant $p$-forms, their derivatives, and the fluxes.

In our conventions, the gravitino variation in eleven dimensions is
\begin{equation}
\delta\Psi_{M}=\hat{\nabla}_{M}\eta-\tfrac{1}{288}\left(\hat{\Gamma}^{NPQR}{}_{M}G_{NPQR}-8G_{MNPQ}\hat{\Gamma}^{NPQ}\right)\eta.
\end{equation}
Before determining the three-dimensional superpotential, we must first identify the correctly normalised three-dimensional gravitino, which in this subsection we denote by $\psi'_{\mu}$. Following, for example, \cite{Coimbra:2012af}, the kinetic terms in the fermion Lagrangian may be diagonalised by redefining the external component of the gravitino as
\begin{equation}
\Psi'_{\mu}=\Psi_{\mu}+\frac{1}{d-2}\hat{\Gamma}_{\mu}\hat{\Gamma}^{m}\Psi_{m}\equiv\Psi_{\mu}+\hat{\Gamma}_{\mu}\hat{\Gamma}^{m}\Psi_{m},
\end{equation}
where $d=3$ in our case. The variation $\delta\psi_{\mu}'$ of the three-dimensional gravitino is then defined by
\begin{equation}
\delta\Psi'_{\mu}=\sqrt{2}\ee^{\Delta/2}\delta\psi_{\mu}'\otimes\epsilon_{+}+\dots\label{eq:redefined_gravitino_variation}
\end{equation}
where the ``\ldots '' denote terms that are not singlets under the local $\Gx 2$ structure on $X$.\footnote{Here $\epsilon_{+}$ is the invariant spinor defining the $G$-structure and hence singles out a particular three-dimensional $N=1$ supersymmetry. The singlet projection does not by itself guarantee that the corresponding gravitino is parametrically lighter than the Kaluza--Klein scale.} Furthermore, $\delta\psi'_{\mu}$ should be independent of the internal coordinates, since it is a three-dimensional field. Since $\Vert\epsilon_{+}\Vert^{2}=1$, we can isolate the $\Gx 2$-singlet component of \eqref{eq:redefined_gravitino_variation} by contracting with $\epsilon^{\transpose}_{+}$ and integrating over $X$ with the warped volume measure. We denote this operation by the projector 
\begin{equation}
\Pi_{+}[\bullet]=\mathcal{V}^{-1}\int\ee^{9\Delta}\vol_{8}\epsilon^{\transpose}_{+}\bullet.
\end{equation}
Applying this to the gravitino variation gives
\begin{equation}
\frac{1}{\sqrt{2}}\Pi_{+}\bigl[\ee^{-\Delta/2}\delta\Psi'_{\mu}\bigr]=\delta\psi_{\mu}'\equiv\nabla_{\mu}\psi-\mathcal{W}\gamma_{\mu}\psi,\label{eq:grav_W}
\end{equation}
where we have used \eqref{eq:3d_shift} to relate the three-dimensional gravitino variation to the real superpotential $\mathcal{W}$.

In order to compute $\delta\Psi'_{\mu}$, we need the external and internal components of the eleven-dimensional gravitino variation. Using the warped-product ansatz and the decomposition of the flux, these are given by \cite{Martelli:2003ki}
\begin{align}
\begin{split}\delta\Psi_{\mu} & =\nabla_{\mu}\eta-\tfrac{1}{2}(\gamma_{\mu}\otimes\partial_{m}\Delta\gamma^{m}\gamma_{9})\eta-\tfrac{1}{288}G_{npqr}\ee^{-3\Delta}(\gamma_{\mu}\otimes\gamma^{npqr}\gamma_{9})\eta\\
 & \eqspace+\tfrac{1}{6}\ee^{-3\Delta}(\gamma_{\mu}\otimes g_{q}\gamma^{q})\eta,
\end{split}
\\
\begin{split}\delta\Psi_{m} & =\nabla_{m}\eta+\tfrac{1}{2}(\id_{2}\otimes\gamma_{m}{}^{a}\partial_{a}\Delta)\eta-\tfrac{1}{288}\ee^{-3\Delta}\bigl((\id_{2}\otimes G_{npqr}\gamma^{npqr}{}_{m})-8(\id_{2}\otimes G_{mnpq}\gamma^{npq})\bigr)\eta\\
 & \eqspace+\tfrac{1}{12}\ee^{-3\Delta}(\id_{2}\otimes\gamma_{m}{}^{r}g_{r}\gamma_{9})\eta-\tfrac{1}{6}\ee^{-3\Delta}(\id_{2}\otimes\gamma_{9})g_{m}\eta,
\end{split}
\end{align}
where $G_{\mu\nu\rho m}=\epsilon_{\mu\nu\rho}g_{m}$. Substituting the spinor ansatz \eqref{eq:spinor_ansatz} into the projected variation and combining the contributions from $\delta\Psi_{\mu}$ and $\delta\Psi_{m}$, we obtain
\begin{align}
 & \frac{1}{\sqrt{2}}\Pi_{+}\bigl[\ee^{-\Delta/2}\delta\Psi'_{\mu}\bigr]\nonumber \\
 & =\nabla_{\mu}\psi+\gamma_{\mu}\psi\,\mathcal{V}^{-1}\int\ee^{9\Delta}\vol_{8}\left(\epsilon^{\transpose}_{+}\gamma_{9}\gamma^{m}\nabla_{m}\epsilon_{+}+\tfrac{1}{96}\ee^{-3\Delta}G_{mnpq}\star Y^{mnpq}-\tfrac{1}{4}\ee^{-3\Delta}g_{m}\bar{K}^{m}\right).
\end{align}
Comparing with the three-dimensional gravitino variation \eqref{eq:grav_W}, we read off
\begin{align}
\mathcal{W} & =\mathcal{V}^{-1}\int\ee^{9\Delta}\vol_{8}\left(-\epsilon^{\transpose}_{+}\gamma_{9}\gamma^{m}\nabla_{m}\epsilon_{+}-\tfrac{1}{96}\ee^{-3\Delta}G_{mnpq}\star Y^{mnpq}+\tfrac{1}{4}\ee^{-3\Delta}g_{m}\bar{K}^{m}\right)\nonumber \\
 & =-\mathcal{V}^{-1}\int\ee^{9\Delta}\vol_{8}\left(\epsilon^{\transpose}_{-}\gamma^{m}\nabla_{m}\epsilon_{+}+\tfrac{1}{4}\vol^{-1}_{8}Y\wedge F-\tfrac{1}{4}\vol^{-1}_{8}\star f\wedge\bar{K}\right),\label{eq:superpotential_intermediate}
\end{align}
where we have used \eqref{eq:flux_warp} to write the $G$-flux in terms of $F$ and $f$. Finally, the rescalings \eqref{eq:rescalings} required to reach three-dimensional Einstein frame imply that $\mathcal{W}$ picks up an additional factor of $\mathcal{V}^{-1}$, i.e.~$\mathcal{W}\rightarrow\mathcal{V}^{-1}\mathcal{W}$, and so our expression for the superpotential is
\begin{equation}
\mathcal{W}=-\frac{1}{\mathcal{V}^{2}}\int\ee^{9\Delta}(\vol_{8}\epsilon^{\transpose}_{-}\gamma^{n}\nabla_{n}\epsilon_{+}+\tfrac{1}{4}F\wedge Y-\tfrac{1}{4}\star f\wedge\bar{K}).\label{eq:superpotential_variation}
\end{equation}

\subsection{Dimensional reduction of the gravitino action}

In this subsection, following the approach of \cite{hep-th/0408121,House:2004pm}, we dimensionally reduce the eleven-dimensional gravitino action \eqref{eq:11d_fermionic_action} in order to identify the mass term of the three-dimensional gravitino. The coefficient of this term determines the superpotential $\mathcal{W}$. We will show that the resulting expression agrees with the one obtained in the previous subsection.

\subsubsection{Three $\Gamma$ term}

We first consider the term in \eqref{eq:11d_fermionic_action} containing three eleven-dimensional gamma matrices. This gives a contribution to the gravitino mass term in three dimensions when the covariant derivative acts along the internal space while the remaining gamma-matrix indices lie along the external directions. Using the metric ansatz \eqref{eq:11_metric} and the gamma-matrix decomposition \eqref{eq:gamma_decomposition}, the derivative of the gravitino can be written in terms of the unwarped internal metric $g$ as 
\begin{equation}
\hat{\nabla}_{n}\Psi_{\rho}=\nabla_{n}\Psi_{\rho}+\tfrac{1}{2}(\id_{2}\otimes\gamma_{n}{}^{p}\partial_{p}\Delta)\Psi_{\rho}-\partial_{n}\Delta\Psi_{\rho}.
\end{equation}
Substituting this into the action, the three-$\Gamma$ term becomes 
\begin{align}
-\frac{1}{\kappa^{2}}\int\dd^{11}X\sqrt{-\hat{g}}\,\bar{\Psi}_{M}\hat{\Gamma}^{MNP}\hat{\nabla}_{N}\hat{\Psi}_{P} & =\frac{1}{\kappa^{2}}\int\dd^{3}y\sqrt{-h}\,\dd^{8}x\sqrt{g}\,\ee^{11\Delta}\bar{\Psi}_{\mu}\Gamma^{\mu\rho}\Gamma^{n}\hat{\nabla}_{n}\Psi_{\rho}\nonumber \\
 & =\frac{2}{\kappa^{2}}\int\dd^{3}y\sqrt{-h}\,\bar{\psi}_{\mu}\gamma^{\mu\nu}\psi_{\nu}\int\dd^{8}x\sqrt{g}\,\ee^{9\Delta}\epsilon^{\transpose}_{-}\gamma^{n}\nabla_{n}\epsilon_{+}.
\end{align}

\subsubsection{Six $\Gamma$ term}

Next we consider the term in \eqref{eq:11d_fermionic_action} containing six eleven-dimensional gamma matrices. This gives a contribution to the gravitino mass term when both gravitino indices lie along the external directions. The remaining indices must contract with the four-form flux. From the decomposition \eqref{eq:flux_warp}, the flux may have either three external legs or none. If three legs were external, the gamma matrix would contain at least five external indices, which vanishes by antisymmetry in three dimensions. The flux must therefore be purely internal in this term. The reduction then gives
\begin{align}
\frac{1}{\kappa^{2}}\int\dd^{11}X\sqrt{-\hat{g}}\,\tfrac{1}{96}G_{PQRS}\bar{\Psi}_{M}\hat{\Gamma}^{MPQRSN}\Psi_{N} & =\frac{2}{\kappa^{2}}\int\dd^{3}y\sqrt{-h}\,\bar{\psi}_{\mu}\gamma^{\mu\nu}\psi_{\nu}\nonumber \\
 & \eqspace\phantom{\frac{2}{\kappa^{2}}}\times\int\dd^{8}x\sqrt{g}\,\ee^{9\Delta}\tfrac{1}{96}F_{pqrs}\epsilon^{\transpose}_{-}\gamma^{pqrs}\epsilon_{+}\nonumber \\
 & =\frac{2}{\kappa^{2}}\int\dd^{3}y\sqrt{-h}\,\bar{\psi}_{\mu}\gamma^{\mu\nu}\psi_{\nu}\int\ee^{9\Delta}\tfrac{1}{4}F\wedge Y.
\end{align}

\subsubsection{Two $\Gamma$ term}

Finally we consider the term in \eqref{eq:11d_fermionic_action} containing two eleven-dimensional gamma matrices. This gives a contribution to the gravitino mass term when both gravitino indices are external. In this case the relevant flux component has three external legs and one internal leg. The resulting expression contains a single external gamma matrix. However, as we review in Appendix \ref{app:conventions}, in three dimensions one has the identities $\gamma^{\sigma}=\tfrac{1}{2}\epsilon^{\sigma\mu\nu}\gamma_{\mu\nu}$ and $\gamma_{\mu\nu}=-\epsilon_{\mu\nu\rho}\gamma^{\rho}$, which convert the expression into a gravitino mass term. The reduction then yields
\begin{align}
\frac{1}{\kappa^{2}}\int\dd^{11}X\sqrt{-\hat{g}}\,\tfrac{1}{8}G_{MNPQ}\bar{\Psi}^{M}\hat{\Gamma}^{NP}\Psi^{Q} & =\frac{1}{\kappa^{2}}\int\dd^{3}y\sqrt{-h}\,\tfrac{1}{2}\epsilon_{\mu\theta\nu}\bar{\psi}^{\mu}\gamma^{\nu}\psi^{\theta}\int\dd^{8}x\sqrt{g}\,\ee^{9\Delta}f_{m}\bar{K}^{m}\nonumber \\
 & =\frac{2}{\kappa^{2}}\int\dd^{3}y\sqrt{-h}\,\bar{\psi}_{\mu}\gamma^{\mu\sigma}\psi_{\sigma}\int\dd^{8}x\sqrt{g}\,\ee^{9\Delta}(-\tfrac{1}{4}f_{m}\bar{K}^{m})\nonumber \\
 & =\frac{2}{\kappa^{2}}\int\dd^{3}y\sqrt{-h}\,\bar{\psi}_{\mu}\gamma^{\mu\nu}\psi_{\nu}\int\ee^{9\Delta}(-\tfrac{1}{4}\star f\wedge\bar{K}),
\end{align}
where we used \eqref{eq:flux_warp} to write
\begin{equation}
G_{\mu\nu\rho m}=\ee^{3\Delta}(\vol_{3})_{\mu\nu\rho}f_{m},\qquad(\vol_{3})_{\mu\nu\rho}=\epsilon_{\mu\nu\rho}.
\end{equation}

\subsubsection{Result}

Adding the three contributions, the mass-like terms in the three-dimensional gravitino action are
\begin{equation}
\frac{2}{\kappa^{2}}\int\dd^{3}y\sqrt{-h}\,\bar{\psi}_{\mu}\gamma^{\mu\nu}\psi_{\nu}\int\ee^{9\Delta}(\vol_{8}\epsilon^{\transpose}_{-}\gamma^{n}\nabla_{n}\epsilon_{+}+\tfrac{1}{4}F\wedge Y-\tfrac{1}{4}\star f\wedge\bar{K}).
\end{equation}
Comparing with the three-dimensional gravitino action \eqref{eq:3d_fermion_action}, setting $\kappa=1$, and including the rescalings \eqref{eq:rescalings} required to reach Einstein frame with a canonical gravitino kinetic term, we obtain
\begin{equation}
\mathcal{W} =-\frac{1}{\mathcal{V}^{2}}\int\ee^{9\Delta}(\vol_{8}\epsilon^{\transpose}_{-}\gamma^{n}\nabla_{n}\epsilon_{+}+\tfrac{1}{4}F\wedge Y-\tfrac{1}{4}\star f\wedge\bar{K}).\label{eq:superpotential_action}
\end{equation}
This agrees with the superpotential derived from the gravitino variation in \eqref{eq:superpotential_variation}.

\subsection{The \texorpdfstring{$N=1$}{N = 1} superpotential}

At this point, we have an expression for the superpotential in terms of the fluxes $F$ and $f$, and the derivative of the invariant spinor $\epsilon_{+}$. We now rewrite this expression in terms of the invariant forms \eqref{eq:invariant_forms} and their derivatives. Note that this must be done without imposing the Killing spinor equations as this would put the superpotential on-shell, whereas it should be an \emph{off-shell} quantity. As a preliminary check, however, it is useful to evaluate our expression on shell.

Using the first Killing spinor equation \eqref{eq:KSE-1}, we find
\begin{align}
\epsilon^{\transpose}_{-}\gamma^{m}\nabla_{m}\epsilon_{+} & =-\tfrac{1}{24}F_{mpqr}\epsilon^{\transpose}_{-}\gamma^{m}\gamma^{pqr}\epsilon_{+}+\tfrac{1}{4}f_{n}\epsilon^{\transpose}_{-}\gamma^{m}\gamma^{n}{}_{m}\epsilon_{-}-m\epsilon^{\transpose}_{-}\gamma^{m}\gamma_{m}\epsilon_{-}\nonumber \\
 & =-\star Y\lrcorner F+\tfrac{7}{4}\bar{K}\lrcorner f-8m\nonumber \\
 & =\vol^{-1}_{8}(-Y\wedge F+\tfrac{7}{4}\star\bar{K}\wedge f-8m\vol_{8}).\label{eq:fierz_result}
\end{align}
Substituting this into the expression for the superpotential from \eqref{eq:superpotential_intermediate} or \eqref{eq:superpotential_action} gives
\begin{equation}
\langle\mathcal{W}\rangle=\frac{1}{\mathcal{V}^{2}}\int\ee^{9\Delta}(\tfrac{3}{4}Y\wedge F-\tfrac{3}{2}\star\bar{K}\wedge f+8m\vol_{8}).\label{eq:superpotential_variation-1}
\end{equation}
As we discuss in Appendix \ref{app:bilinears} and give in \eqref{eq:ext_8_1}, one of the algebraic conditions implied by the second Killing spinor equation \eqref{eq:KSE-2} is
\begin{equation}
Y\wedge F-2\star\bar{K}\wedge f+12m\vol_{8}=0,
\end{equation}
which implies that \eqref{eq:superpotential_variation-1} simplifies to $\langle\mathcal{W}\rangle=-m/\mathcal{V}$ on-shell. This agrees with the condition \eqref{eq:3d_susy} for a supersymmetric vacuum in three dimensions, together with the relation between $L_{\text{AdS}}$ and $m$ given in \eqref{eq:rescaled}.

We now return to the off-shell computation. As we show in detail in Appendix \ref{sec:Computing-the-superpotential}, using $\epsilon^{\transpose}_{+}\epsilon_{+}=1$, $\epsilon^{\transpose}_{-}\gamma^{m}\epsilon_{+}=0$ and the Fierz identity
\begin{equation}
2^{4}\epsilon_{+}\epsilon^{\transpose}_{+}=\id_{16}+\gamma^{m}\bar{K}_{m}+\tfrac{1}{4!}\gamma^{mnpq}Y_{mnpq}-\tfrac{1}{5!}\gamma^{mnpqr}\star\bar{\phi}_{mnpqr}+\gamma_{9}\sin\zeta,
\end{equation}
we may rewrite the spinorial term as 
\begin{align}
2^{4}\epsilon^{\transpose}_{-}\gamma^{r}\nabla_{r}\epsilon_{+} & =2^{4}(\epsilon^{\transpose}_{-}\gamma^{r}\nabla_{r}\epsilon_{+})(\epsilon^{\transpose}_{+}\epsilon_{+})=2^{4}\epsilon^{\transpose}_{-}\gamma^{r}\nabla_{r}(\epsilon_{+}\epsilon^{\transpose}_{+})\epsilon_{+}-2^{4}\cancel{\epsilon^{\transpose}_{-}\gamma^{m}\epsilon_{+}}\nabla_{m}\epsilon^{\transpose}_{+}\epsilon_{+}\nonumber \\
 & =\sin\zeta\,\nabla_{a}\bar{K}^{a}+\tfrac{1}{3!}\bar{\phi}_{npq}\nabla_{m}Y^{mnpq}-\tfrac{1}{4!}\star Y_{npqr}\nabla_{m}\star\bar{\phi}^{mnpqr}-\bar{K}^{a}\nabla_{a}\sin\zeta\nonumber \\
 & =\vol^{-1}_{8}(-8\sin\zeta\,\dd\star\bar{K}-2\bar{\phi}\wedge\dd\star Y+6\star\bar{K}\wedge\dd\sin\zeta),\label{eq:Fierz_differential}
\end{align}
where we have expressed the result using the spinor bilinears collected in Appendix \ref{app:G2-bilinears}. The off-shell superpotential becomes
\begin{equation}
	\mathcal{W} =-\frac{1}{8\mathcal{V}^{2}}\int\ee^{9\Delta}\left(-4\sin\zeta\,\dd\star\bar{K}-\bar{\phi}\wedge\dd\star Y+3\star\bar{K}\wedge\dd\sin\zeta+2\,Y\wedge F-2\,\star f\wedge\bar{K}\right).\label{eq:superpotential_barred}
\end{equation}
This expression involves only the globally defined bilinears $\bar{K}$, $\bar{\phi}$, $Y$ and $\sin\zeta$, so it remains smooth in the $\Spin7$ limit, $\sin\zeta=\pm1$. It defines a real functional on the space of local $\Gx2$ structures and fluxes on $X$, depending only on the invariant forms and their exterior derivatives. Since we derived it without imposing the Killing spinor equations, it remains valid off-shell.

The supersymmetry conditions \eqref{eq:sufficient_1}--\eqref{eq:sufficient_6}, however, are written in terms of the canonically normalised $\Gx2$ forms $K$ and $\phi$. Rewriting the superpotential in terms of these gives
\begin{equation}
	\mathcal{W}=\frac{1}{8\mathcal{V}^{2}}\int\ee^{9\Delta}\Bigl(\tfrac{4}{7}\dd\zeta\wedge\phi\wedge\psi+\cos\zeta\,\phi\wedge K\wedge\dd\phi-2\,Y\wedge F+2\cos\zeta\star f\wedge K\Bigr).\label{eq:superpotential}
\end{equation}
We will use this form of the superpotential to study its critical points and connect to the supersymmetry conditions.

\subsubsection{Reduction to the $\protect\Spin 7$ superpotential}

As a consistency check, it is instructive to compare the off-shell superpotential with the known expression for the M-theory superpotential on an eight-manifold with a \emph{global} $\Spin 7$ structure. This superpotential was first conjectured by Gukov in \cite{Gukov:1999gr} and later shown in \cite{Acharya:2002vs,Becker:2002jj} to reproduce the supersymmetry conditions of \cite{Becker:2000jc}. For its appearance in the finite-dimensional three-dimensional effective action, see~\cite{1307.5858}.

A global $\Spin 7$ structure is defined by a single nowhere-vanishing Majorana--Weyl spinor, which in the notation of \eqref{eq:spinor_ansatz} we may take to be $\xi_{+}$, with $\xi_{-}=0$ and hence $\sin\zeta=1$. In this limit, of the globally defined bilinears, only the four-form $Y$ survives, which is then identified with the self-dual Cayley form for the $\Spin 7$ structure. Our expression for the superpotential then reduces to
\begin{equation}
\left.\mathcal{W}\right|_{\Spin 7}=-\frac{1}{4\mathcal{V}^{2}}\int\ee^{9\Delta}Y\wedge F,\label{eq:spin7_superpotential}
\end{equation}
in agreement with \cite{Gukov:1999gr} up to the inclusion of warping.\footnote{See also \cite{Becker:2003wb} for the matching between the superpotential and the corresponding three-dimensional $N=1$ supergravity on a $\Spin 7$ holonomy manifold.} 

At first sight, this simpler functional appears not to capture the full supersymmetry conditions of the $\Spin 7$ case. In particular, as discussed in \cite{Acharya:2002vs,Martelli:2003ki}, supersymmetry allows an electric flux of the form $f=3\,\dd\Delta$, which clearly cannot be determined from \eqref{eq:spin7_superpotential} alone. The resolution is that \eqref{eq:spin7_superpotential} is obtained by restricting the general off-shell functional to the $\Spin 7$ locus. To recover the full supersymmetry conditions, one must instead vary the general superpotential \eqref{eq:superpotential} and only then specialise to the $\Spin 7$ limit. We will return to this point at the end of the next section.

\section{Supersymmetry conditions from the superpotential\label{sec:Supersymmetry-conditions-from}}

In this section we show that supersymmetric Minkowski or AdS solutions arise as critical points of the real superpotential $\mathcal{W}$. More precisely, we will show that requiring the superpotential and its variation to satisfy 
\begin{equation}
\langle\delta\mathcal{W}\rangle=0,\qquad\langle\mathcal{W}\rangle=-\frac{m}{\mathcal{V}},
\end{equation}
reproduces all of the supersymmetry conditions \eqref{eq:sufficient_1}--\eqref{eq:sufficient_6} for the underlying eleven-dimensional solution.

As reviewed in Section \ref{subsec:Three-dimensional--supergravity}, the three-dimensional effective theory is minimal supergravity coupled to matter. The bosonic matter sector is a sigma model whose scalar fields locally parametrise a Riemannian target space.\footnote{See also \cite[Section 4]{deIaOssa:2019cci} for related discussion in the context of the heterotic string compactified on a seven-manifold with a $\Gx 2$ structure.} Supersymmetric vacua therefore require the vanishing of the variation of the superpotential with respect to these scalar fields. In our compactification the scalar fields arise from deformations of the eight-dimensional geometry and fluxes. 

The eleven-dimensional background is specified by the fields $(\Delta,\zeta,K,\phi,F,f)$ on the internal manifold $X$, while variations of these fields correspond to scalar fluctuations in the three-dimensional theory. To analyse the critical points of $\mathcal{W}$, we must therefore consider the most general variations of these quantities. It is convenient to organise these variations according to their transformation properties under the local $\Gx 2$ structure. 

The local $\Gx 2$ structure is defined by the pair $(K,\phi)$. Breaking $\SO 8\to\SO 7\to\Gx 2$, a one-form and three-form admit the following decomposition into $\Gx 2$ representations:
\begin{align}
\text{one-form}\in\rep 8_{\charge} & \to\rep 1 \oplus\rep 7\to\rep 1\oplus\rep 7,\\
\text{three-form}\in\rep{56}_{\charge} & \to\rep{21}\oplus\rep{35}\to\rep 7\oplus\rep{14}\oplus\rep 1\oplus\rep 7\oplus\rep{27}.
\end{align}
Hence, we can parametrise variations of $K$ and $\phi$ as
\begin{align}
\delta K & =\mu^{\rep 1}_{0}K+\mu^{\rep 7}_{1},\label{eq:K_variations}\\
\delta\phi & =K\wedge(\mu^{\rep 7}_{2}+\mu^{\rep{14}}_{2})+\tilde{\mu}^{\rep 1}_{0}\phi+\tilde{\mu}^{\rep 7}_{1}\lrcorner\psi+\mu^{\rep{27}}_{3},\label{eq:phi_variations}
\end{align}
where our notation is that $\mu^{\rep r}_{p}$ is a $p$-form transforming in the $\Gx 2$ representation $\rep r$, and $\lrcorner$ denotes the interior product. As we discuss in Appendix \ref{subsec:Eight-dimensions}, since we assume the variations of $(K,\phi)$ give another local $\Gx 2$ structure, the $\delta\phi$ variation also fixes $\delta\psi$.

We must also specify which flux degrees of freedom are held fixed when varying the geometry. We use the internal three-form potential $A$ and its dual six-form potential $\tilde A$, with local field strengths
\begin{equation}
	\mathcal F\equiv\dd A=\ee^{3\Delta}F,\qquad
	\mathcal H\equiv\dd\tilde A-\tfrac12 A\wedge\dd A
	=\ee^{6\Delta}\star f,\label{eq:potentials}
\end{equation}
which satisfy
\begin{equation}
	\dd\mathcal F=0,\qquad
	\dd\mathcal H=-\tfrac12\mathcal F\wedge\mathcal F.
\end{equation}
In the three-form description of the eleven-dimensional flux, the second relation is an equation of motion, whereas in the dual-potential description it follows from the definition of $\mathcal H$. We will therefore vary within this constrained space of fluxes, without imposing the Killing spinor equations or the remaining field equations. Given this, $F$ and $f$ are not independent variables: they are determined by the potentials and the geometry through \eqref{eq:potentials}. When varying the geometry, we hold the potentials fixed within a given flux sector. Thus, for example, at fixed $\Delta$ a metric variation keeps $\star f$ fixed while $f$ varies through the Hodge star.

For what follows, it will be useful to decompose the fluxes into $\SO 7$ representations with respect to the vector $K$ as
\begin{equation}
F=F_{4}+F_{3}\wedge K,\qquad f=f_{1}+f_{0}K,\label{eq:flux_so7}
\end{equation}
where $\imath_{K}F_{4}=\imath_{K}f_{1}=0$. Our conventions for the Hodge star then imply $\star F=-\star_{7}F_{4}\wedge K-\star_{7}F_{3}$ and $\star f=\star_{7}f_{1}\wedge K-f_{0}\star_{7}1$.

Putting this together, supersymmetric vacua should correspond to critical points of $\mathcal{W}$ satisfying
\begin{equation}
\langle\delta\mathcal{W}\rangle=0,\qquad\langle\mathcal{W}\rangle=-\frac{m}{\mathcal{V}},
\end{equation}
where the variation $\delta\mathcal{W}$ includes all admissible scalar deformations.\footnote{The displayed list also includes a formal $\mu^{\rep{14}}_{2}$ direction. It is not tangent to the space of $\Gx2$ structures and is excluded from the physical variations; below we check that its formal variation imposes no additional equation once \eqref{eq:dK} holds.}
\begin{equation}
\bigl(\delta\Delta,\delta\zeta,\mu^{\rep 1}_{0},\tilde{\mu}^{\rep 1}_{0},\mu^{\rep 7}_{1},\mu^{\rep 7}_{2},\tilde{\mu}^{\rep 7}_{1},\mu^{\rep{14}}_{2},\mu^{\rep{27}}_{3},\delta A,\delta\tilde{A}\bigr).\label{eq:variations}
\end{equation}
In the remainder of this section, we will evaluate these variations and show that the resulting conditions reproduce \emph{all} of the supersymmetry constraints \eqref{eq:sufficient_1}--\eqref{eq:sufficient_6} for the eleven-dimensional solution. Note that among these variations, only $\delta\Delta$ and the singlets $\mu^{\rep 1}_{0}$ and $\tilde{\mu}^{\rep 1}_{0}$ can change the volume of the internal manifold $X$. Thanks to this, the $\mathcal{V}^{-2}$ prefactor in $\mathcal{W}$ can be ignored for the other variations. We now vary the superpotential with respect to each independent scalar deformation and finally impose that the superpotential itself evaluates to $-m/\mathcal{V}$.

\subsection{Critical points of the superpotential}

\subsubsection{Variation w.r.t.~$\tilde{A}$}

We begin with the variation of the six-form potential $\tilde{A}$ for $\star f$. Starting from the superpotential \eqref{eq:superpotential}, varying with respect to $\tilde{A}$ gives
\begin{align}
\delta_{\tilde{A}}\mathcal{W} & =\frac{1}{4\mathcal{V}^{2}}\int\ee^{9\Delta}\Bigl(\cos\zeta\,\delta_{\tilde{A}}\star f\wedge K\Bigr)
  =\frac{1}{4\mathcal{V}^{2}}\int\ee^{3\Delta}\Bigl(\cos\zeta\,\dd\delta\tilde{A}\wedge K\Bigr)\nonumber \\
 & =-\frac{1}{4\mathcal{V}^{2}}\int\dd(\ee^{3\Delta}\cos\zeta\,K)\wedge\delta\tilde{A},
\end{align}
where in the final step we integrated by parts. This vanishes for all $\delta\tilde{A}$ provided
\begin{equation}
\dd(\ee^{3\Delta}\cos\zeta\,K)=0.\label{eq:dK}
\end{equation}
This is precisely Equation \eqref{eq:sufficient_1} of the necessary and sufficient supersymmetry conditions. Comparing with the torsion classes in \eqref{eq:torsion1} for a $\Gx 2$ structure in eight dimensions, this condition fixes
\begin{equation}
\tilde{\beta}_{\rep 7}=3\dd_{7}\Delta-\tan\zeta\,\dd_{7}\zeta,\qquad\gamma_{\rep 7}=0,\qquad\gamma_{\rep{14}}=0,\label{eq:K_tor}
\end{equation}
where we have decomposed the exterior derivative on $X$ as $\dd=\dd_{7}+K\wedge\partial_{K}$.

\subsubsection{Variation w.r.t.~$A$}

We next vary the three-form potential $A$, keeping $\tilde A$, the metric and $\Delta$ fixed. Since $A$ enters both field strengths in \eqref{eq:potentials}, one has
\begin{align}
\delta_A\mathcal F&=\dd\delta A,\\
\delta_A\mathcal H&=-\tfrac12\delta A\wedge\dd A-\tfrac12 A\wedge\dd\delta A
=-\delta A\wedge\mathcal F+\tfrac12\dd(A\wedge\delta A).
\end{align}
Writing $B=\ee^{3\Delta}\cos\zeta\,K$, the variation becomes
\begin{align}
\delta_A\mathcal W
&=\frac{1}{4\mathcal V^2}\int\left[-\ee^{6\Delta}Y\wedge\dd\delta A
+\left(-\delta A\wedge\mathcal F+\tfrac12\dd(A\wedge\delta A)\right)\wedge B\right]\nonumber\\
&=\frac{1}{4\mathcal V^2}\int\left[\dd(\ee^{6\Delta}Y)+B\wedge\mathcal F-\tfrac12\dd B\wedge A\right]\wedge\delta A.
\end{align}
The third term is proportional to the $\tilde A$ critical-point equation $\dd B=0$ in \eqref{eq:dK}.\footnote{Equivalently, the flux variations can be expressed using $\delta\tilde A+\tfrac12 A\wedge\delta A$ and $\delta A$ as a basis.}  After imposing the $\tilde A$ equation, the variation vanishes for all $\delta A$ provided
\begin{equation}
\ee^{-6\Delta}\dd(\ee^{6\Delta}Y)=-\cos\zeta\,K\wedge F,
\end{equation}
which is Equation \eqref{eq:dY} of the supersymmetry conditions in Appendix \ref{app:bilinears}.\footnote{This was also derived in \cite[Equation 4.15]{Martelli:2003ki}.} Comparing with the torsion classes in \eqref{eq:torsion1}--\eqref{eq:torsion3}, this condition fixes
\begin{align}
(-6\partial_{K}\Delta-4\tilde{\alpha}_{\rep 1}+\sin\zeta\,\alpha_{\rep 1})\psi & =-\cos\zeta\,\pi_{\rep 1}(F_{4}), & \eta_{\rep{14}} & =0,\label{eq:14_Y}\\
\frac{1}{\cos\zeta}\dd_{7}\zeta\wedge\phi-\tfrac{3}{2}\sin\zeta\,\dd_{7}\Delta\wedge\phi-\star_{7}\tilde{\delta}_{\rep 7} & =-\cos\zeta\,\pi_{\rep 7}(F_{4}), & \beta_{\rep 7} & =-\tfrac{3}{2}\dd_{7}\Delta,\label{eq:beta7}\\
\star_{7}\tilde{\delta}_{\rep{27}}+\sin\zeta\star_7\delta_{\rep{27}} & =-\cos\zeta\,\pi_{\rep{27}}(F_{4}),\label{eq:27_2}
\end{align}
where $\pi_{\boldsymbol{r}}$ denotes the projector onto the $\Gx 2$ representation $\rep r$.

\subsubsection{Variation w.r.t~$\zeta$}

We next vary  the scalar function $\zeta$. This gives
\begin{align}
\delta_{\zeta}\mathcal{W} & =\frac{1}{8\mathcal{V}^{2}}\int\ee^{9\Delta}\Bigl(\tfrac{4}{7}\dd\delta\zeta\wedge\phi\wedge\psi-\sin\zeta\,\delta\zeta\,\phi\wedge K\wedge\dd\phi-2\cos\zeta\,\delta\zeta\,\phi\wedge K\wedge F-2\sin\zeta\,\delta\zeta\star f\wedge K\Bigr)\nonumber \\
 & =-\frac{1}{8\mathcal{V}^{2}}\int\ee^{9\Delta}\Bigl(\tfrac{4}{7}\ee^{-9\Delta}\dd(\ee^{9\Delta}\phi\wedge\psi)+\bigl(\phi\wedge(\sin\zeta\,\dd\phi+2\cos\zeta\,F)+2\sin\zeta\star f\bigr)\wedge K\Bigr)\delta\zeta
\end{align}
For this to vanish for all $\delta\zeta$, we must have
\begin{equation}
0=\tfrac{36}{7}\dd\Delta\wedge\phi\wedge\psi+\tfrac{4}{7}\dd(\phi\wedge\psi)+\sin\zeta\,\phi\wedge K\wedge\dd\phi+2\cos\zeta\,\phi\wedge K\wedge F+2\sin\zeta\star f\wedge K.
\end{equation}
Substituting the torsion-class decompositions \eqref{eq:torsion2} and \eqref{eq:torsion3} for $\dd\phi$ and $\dd\psi$, we find
\begin{equation}
0=\left(\tfrac{36}{7}\partial_{K}\Delta+4\tilde{\alpha}_{\rep 1}-\sin\zeta\,\alpha_{\rep 1}+\tfrac{2}{7}\sin\zeta\,f_{0}\right)\psi-2\cos\zeta\,\pi_{\rep 1}(F_{4}).\label{eq:xi_1}
\end{equation}
This equation relates the singlet classes $\alpha_{\rep 1}$ and $\tilde{\alpha}_{\rep 1}$ to $\partial_{K}\Delta$, the singlet $\pi_{\rep 1}(F_{4})\propto\psi$, and the component of $f$ along $K$.

\subsubsection{Variation w.r.t.~$\tilde{\mu}^{\protect\rep 7}_{1}$}

Next we vary the $\rep 7$ component of $\phi$ orthogonal to $K$. Writing this variation as $\delta_{\rep 7}\phi=\tilde{\mu}^{\rep 7}_{1}\lrcorner\psi$, we find
\begin{align}
\delta_{\rep 7}\mathcal{W} & =\frac{1}{8\mathcal{V}^{2}}\int\ee^{9\Delta}\Bigl(\tfrac{4}{7}\dd\zeta\wedge(\delta_{\rep 7}\phi\wedge\psi+\phi\wedge\delta_{\rep 7}\psi)+\cos\zeta(\delta_{\rep 7}\phi\wedge K\wedge\dd\phi+\phi\wedge K\wedge\dd\delta_{\rep 7}\phi)-2\delta_{\rep 7}Y\wedge F\Bigr)\nonumber \\
 & =-\frac{1}{8\mathcal{V}^{2}}\int\ee^{9\Delta}\left(\cos\zeta\,K\wedge\dd\phi+\ee^{-9\Delta}\dd(\ee^{9\Delta}\cos\zeta\,\phi\wedge K)-2K\wedge(\star F+\sin\zeta\,F)\right)\wedge\delta_{\rep 7}\phi\nonumber \\
 & =-\frac{1}{4\mathcal{V}^{2}}\int\ee^{9\Delta}\left(3\cos\zeta\,\dd\Delta\wedge\phi\wedge K+\cos\zeta\,\dd\phi\wedge K-K\wedge(\star F-\sin\zeta\,F)\right)\wedge\delta_{\rep 7}\phi,
\end{align}
where we used \eqref{eq:dK} to eliminate $\dd K$. Substituting the expression for $\dd\phi$ in terms of the torsion classes, and noting that $\beta_{\rep 7}=-\tfrac{3}{2}\dd_{7}\Delta$ from the derivative of $Y$ in \eqref{eq:beta7}, this vanishes for all $\delta_{\rep 7}\phi$ provided
\begin{equation}
0=-\tfrac{3}{2}\cos\zeta\,\dd_{7}\Delta\wedge\phi+\star_{7}\pi_{\rep 7}(F_{3})-\sin\zeta\,\pi_{\rep 7}(F_{4}).\label{eq:delta_phi_7_2}
\end{equation}
This relates the two $\rep 7$ components of $F$ to the derivative of the warp factor orthogonal to $K$.

\subsubsection{Variation w.r.t.~$\mu^{\protect\rep 7}_{1}$}

Varying the superpotential with respect to $\delta_{\rep 7}K=\mu^{\rep 7}_{1}$ gives
\begin{align}
\delta_{\rep 7}\mathcal{W} & =\frac{1}{8\mathcal{V}^{2}}\int\ee^{9\Delta}\Bigl(\cos\zeta\,\phi\wedge\delta_{\rep 7}K\wedge\dd\phi-2\delta_{\rep 7}Y\wedge F+2\cos\zeta\star f\wedge\delta_{\rep 7}K\Bigr)\nonumber \\
 & =\frac{1}{8\mathcal{V}^{2}}\int\ee^{9\Delta}\Bigl(\cos\zeta\,\phi\wedge\delta_{\rep 7}K\wedge\dd\phi-2\sin\zeta\,\phi\wedge\delta_{\rep 7}K\wedge F+2\cos\zeta\star f\wedge\delta_{\rep 7}K\Bigr)\nonumber \\
 & =\frac{1}{8\mathcal{V}^{2}}\int\ee^{9\Delta}\Bigl(\cos\zeta\,\phi\wedge\dd\phi-2\sin\zeta\,\phi\wedge F+2\cos\zeta\star f\Bigr)\wedge\delta_{\rep 7}K.
\end{align}
Using the expression for $\dd\phi$ in terms of the torsion classes, this vanishes for all $\delta_{\rep 7}K$ provided
\begin{equation}
0=\cos\zeta\,\phi\wedge\tilde{\delta}_{\rep 7}+2\sin\zeta\,\phi\wedge\pi_{\rep 7}(F_{3})-2\cos\zeta\star f_{1}.\label{eq:delta_K_7}
\end{equation}

\subsubsection{Variation w.r.t.~$\mu^{\protect\rep 7}_{2}$}

Varying $\mathcal{W}$ with respect to $\delta_{\rep 7}\phi=K\wedge\mu^{\rep 7}_{2}$ gives
\begin{align}
\delta_{\rep 7}\mathcal{W} & =\frac{1}{8\mathcal{V}^{2}}\int\ee^{9\Delta}\Bigl(\tfrac{4}{7}\dd\zeta\wedge(\delta_{\rep 7}\phi\wedge\psi+\phi\wedge\delta_{\rep 7}\psi)+\cos\zeta(\delta_{\rep 7}\phi\wedge K\wedge\dd\phi+\phi\wedge K\wedge\dd\delta_{\rep 7}\phi)-2\delta_{\rep 7}Y\wedge F\Bigr)\nonumber \\
 & =\frac{1}{8\mathcal{V}^{2}}\int\ee^{9\Delta}\Bigl(\tfrac{4}{7}\dd\zeta\wedge\delta_{\rep 7}\phi\wedge\psi+\tfrac{4}{7}\dd\zeta\wedge\phi\wedge\delta_{\rep 7}\psi-\ee^{-9\Delta}\dd(\ee^{9\Delta}\cos\zeta\,\phi\wedge K)\wedge\delta_{\rep 7}\phi+2\delta_{\rep 7}\psi\wedge F\Bigr)\nonumber \\
 & =\frac{1}{8\mathcal{V}^{2}}\int\ee^{9\Delta}\Bigl(\dd\zeta\wedge\phi+2F\Bigr)\wedge\delta_{\rep 7}\psi,
\end{align}
where we used $\delta_{\rep 7}\phi\wedge\psi=\tfrac{3}{4}\phi\wedge\delta_{\rep 7}\psi$ and $K\wedge\dd K=0$, which follows from \eqref{eq:dK}. Since $\delta_{\rep 7}\psi$ has a component along $K$, this vanishes for all $\delta_{\rep 7}\psi$ provided
\begin{equation}
\pi_{\rep 7}(F_{4})=\tfrac{1}{2}\phi\wedge\dd_{7}\zeta.\label{eq:delta_phi_7}
\end{equation}

\subsubsection{Variation w.r.t.~$\mu^{\protect\rep{14}}_{2}$}

We next vary the $\rep{14}$ component of $\phi$, which we write as $\delta_{\rep{14}}\phi=K\wedge\mu^{\rep{14}}_{2}$. This gives
\begin{align}
\delta_{\rep{14}}\mathcal{W} & =\frac{1}{8\mathcal{V}^{2}}\int\ee^{9\Delta}\Bigl(\cos\zeta\,\delta_{\rep{14}}\phi\wedge K\wedge\dd\phi+\cos\zeta\,\phi\wedge K\wedge\dd\delta_{\rep{14}}\phi-2\delta_{\rep{14}}Y\wedge F\Bigr)\nonumber \\
 & =\frac{1}{8\mathcal{V}^{2}}\int\ee^{9\Delta}\Bigl(-\cos\zeta\,K\wedge\dd\phi-\ee^{-9\Delta}\dd(\ee^{9\Delta}\cos\zeta\,\phi\wedge K)+2\sin\zeta\,K\wedge F\Bigr)\wedge K\wedge\mu^{\rep{14}}_{2}\nonumber \\
 & =\frac{1}{8\mathcal{V}^{2}}\int\ee^{9\Delta}\cos\zeta\,\phi\wedge\dd K\wedge K\wedge\mu^{\rep{14}}_{2}.
\end{align}
Once the critical-point condition \eqref{eq:dK} for $\tilde{A}$ is imposed, $K\wedge\dd K=0$ and this variation vanishes automatically. In fact the $\rep{14}$ direction should not be included among the independent physical variations: pointwise on $X$, the tangent space to a $\Gx2$ structure decomposes as
\begin{equation}
T\bigl(\GL8/\Gx2\bigr)\simeq\rep1\oplus\rep1\oplus3(\rep7)\oplus\rep{27},
\end{equation}
with no $\rep{14}$ component. The form decomposition \eqref{eq:phi_variations} includes $K\wedge\mu^{\rep{14}}_2$, but that direction is not tangent to the space of compatible $\Gx2$ structures. Thus one should not expect to find a further supersymmetry condition.

\subsubsection{Variation w.r.t.~$\mu^{\protect\rep{27}}_{3}$}

Varying $\mathcal{W}$ with respect to $\delta_{\rep{27}}\phi=\mu^{\rep{27}}_{3}$ gives
\begin{align}
\delta_{\rep{27}}\mathcal{W} & =\frac{1}{8\mathcal{V}^{2}}\int\ee^{9\Delta}\Bigl(\cos\zeta\,\delta_{\rep{27}}\phi\wedge K\wedge\dd\phi+\cos\zeta\,\phi\wedge K\wedge\dd\delta_{\rep{27}}\phi-2\delta_{\rep{27}}Y\wedge F\Bigr)\nonumber \\
 & =-\frac{1}{4\mathcal{V}^{2}}\int\ee^{9\Delta}\Bigl(\cos\zeta\,\dd\phi+\star F-\sin\zeta\,F\Bigr)\wedge K\wedge\delta_{\rep{27}}\phi.
\end{align}
Using the expression \eqref{eq:torsion2} for $\dd\phi$ in terms of the torsion classes, this vanishes for all $\delta_{\rep{27}}\phi$ provided
\begin{equation}
0=\cos\zeta\star_{7}\delta_{\rep{27}}-\star_{7}\pi_{\rep{27}}(F_{3})-\sin\zeta\,\pi_{\rep{27}}(F_{4}).\label{eq:27_1}
\end{equation}

\subsubsection{Variations w.r.t.~$\mu^{\protect\rep 1}_{0}$ and $\tilde{\mu}^{\protect\rep 1}_{0}$}

We turn to the singlet variations of $K$ and $\phi$. There are two such singlets, which we denote by $\mu^{\rep 1}_{0}$ and $\tilde{\mu}^{\rep 1}_{0}$ respectively. One linear combination of these leaves $\vol_{8}$ invariant, and so does not affect the $\mathcal{V}^{-2}$ prefactor in $\mathcal{W}$. The other rescales the internal volume and therefore also changes $\mathcal{V}$. Given the variations of $K$ and $\phi$ in \eqref{eq:K_variations} and \eqref{eq:phi_variations}, the volume form defined by the $\Gx 2$ structure on $X$ varies as
\begin{equation}
\delta_{\rep 1}\vol_{8}=(\mu^{\rep 1}_{0}+\tfrac{7}{3}\tilde{\mu}^{\rep 1}_{0})\vol_{8}.
\end{equation}
Hence the combination $\tilde{\mu}^{\rep 1}_{0}=-\tfrac{3}{7}\mu^{\rep 1}_{0}$ leaves the volume invariant.

Varying $\mathcal{W}$ with respect to both singlets gives
\begin{align}
\delta_{\rep 1}\mathcal{W} & =\frac{1}{8\mathcal{V}^{2}}\int\ee^{9\Delta}\Bigl(\tfrac{4}{7}\dd\zeta\wedge\delta_{\rep 1}(\phi\wedge\psi)+\cos\zeta(\delta_{\rep 1}\phi\wedge K\wedge\dd\phi+\phi\wedge\delta_{\rep 1}K\wedge\dd\phi+\phi\wedge K\wedge\dd\delta_{\rep 1}\phi)\nonumber \\
 & \eqspace\phantom{\frac{1}{8\mathcal{V}^{2}}\int\ee^{9\Delta}\Bigl(}-2\delta_{\rep 1}Y\wedge F+2\star f\wedge\delta_{\rep 1}\bar{K}\Bigr)-\frac{2\mathcal{W}}{\mathcal{V}}\int\ee^{9\Delta}\delta_{\rep 1}\vol_{8}\nonumber \\
 & =\frac{1}{8\mathcal{V}^{2}}\int\ee^{9\Delta}\Bigl(\tfrac{4}{3}\tilde{\mu}^{\rep 1}_{0}\partial_{K}\zeta\,K\wedge\phi\wedge\psi-(\mu^{\rep 1}_{0}+2\tilde{\mu}^{\rep 1}_{0})\cos\zeta\,K\wedge\phi\wedge\dd\phi-\tfrac{8}{3}\tilde{\mu}^{\rep 1}_{0}K\wedge F_{3}\wedge\psi\nonumber \\
 & \eqspace\phantom{\frac{1}{8\mathcal{V}^{2}}\int\ee^{9\Delta}\Bigl(}+2K\wedge\phi\wedge\bigl((\mu^{\rep 1}_{0}+\tilde{\mu}^{\rep 1}_{0})\sin\zeta\,F_{4}+\tfrac{1}{7}\mu^{\rep 1}_{0}f_{0}\cos\zeta\,\psi\bigr)\Bigr)-\frac{2\mathcal{W}}{\mathcal{V}}\int\ee^{9\Delta}\vol_{8}(\mu^{\rep 1}_{0}+\tfrac{7}{3}\tilde{\mu}^{\rep 1}_{0})\nonumber \\
 & =\frac{1}{8\mathcal{V}^{2}}\int\ee^{9\Delta}K\wedge\phi\wedge\Bigl(\psi\bigl(\tfrac{4}{3}\tilde{\mu}^{\rep 1}_{0}\partial_{K}\zeta-(\mu^{\rep 1}_{0}+2\tilde{\mu}^{\rep 1}_{0})\cos\zeta\,\alpha_{\rep 1}+\tfrac{2}{7}\mu^{\rep 1}_{0}f_{0}\cos\zeta-\tfrac{16}{7}\mathcal{V}\mathcal{W}(\mu^{\rep 1}_{0}+\tfrac{7}{3}\tilde{\mu}^{\rep 1}_{0})\bigr)\nonumber \\
 & \eqspace\phantom{\frac{1}{8\mathcal{V}^{2}}\int\ee^{9\Delta}K\wedge\phi\wedge\Bigl(}-\tfrac{8}{3}\tilde{\mu}^{\rep 1}_{0}\star_{7}\pi_{\rep 1}(F_{3})+2(\mu^{\rep 1}_{0}+\tilde{\mu}^{\rep 1}_{0})\sin\zeta\,\pi_{\rep 1}(F_{4})\Bigr),
\end{align}
where in the final line we used the decomposition of $F$ and $f$ in \eqref{eq:flux_so7}, together with the torsion-class decomposition \eqref{eq:torsion2} of $\dd\phi$. This vanishes for all $\mu^{\rep 1}_{0}$ and $\tilde{\mu}^{\rep 1}_{0}$ provided the following two conditions hold:
\begin{align}
0 & =(-13\cos\zeta\,\alpha_{\rep 1}+4\partial_{K}\zeta+2f_{0}\cos\zeta-32\mathcal{V}\mathcal{W})\psi+20\sin\zeta\,\pi_{\rep 1}(F_{4})-8\star_{7}\pi_{\rep 1}(F_{3}),\label{eq:singlet_1}\\
0 & =(-\cos\zeta\alpha_{\rep 1}-4\partial_{K}\zeta+2f_{0}\cos\zeta)\psi+8\sin\zeta\,\pi_{\rep 1}(F_{4})+8\star_{7}\pi_{\rep 1}(F_{3}).\label{eq:singlet_2}
\end{align}
We next vary the warp factor at fixed $A$ and $\tilde A$, and combine its equation with the two singlet equations above.

\subsubsection{Variation w.r.t.~$\Delta$}

Finally, we consider variations of the warp factor $\Delta$. As discussed above, the independent flux degrees of freedom are encoded by the potentials $A$ and $\tilde{A}$, rather than by $F$ and $f$ themselves. Thus, when varying $\Delta$ we must hold $A$ and $\tilde{A}$ fixed. From the definitions of the potentials in \eqref{eq:potentials}, we then have 
\begin{equation}
	\delta_{\Delta}F=-3\delta\Delta\,F,\qquad\delta_{\Delta}\star f=-6\delta\Delta\star f.
\end{equation}
The warped volume varies as
\begin{equation}
	\delta_{\Delta}\mathcal{V}=9\int\delta\Delta\,\ee^{9\Delta}\vol_{8}.
\end{equation}

For convenience, define a $\Delta$-independent top-form $\Theta$ as
\begin{equation}
	\Theta=\tfrac{4}{7}\dd\zeta\wedge\phi\wedge\psi+\cos\zeta\,\phi\wedge K\wedge\dd\phi,
\end{equation}
so that the superpotential is 
\begin{equation}
	\mathcal{W}=\frac{1}{8\mathcal{V}^{2}}\int\ee^{9\Delta}(\Theta-2Y\wedge F+2\cos\zeta\star f\wedge K).
\end{equation}
Keeping the potentials fixed, its variation with respect to the warp factor is then
\begin{equation}
	\delta_{\Delta}\mathcal{W}=-\frac{2\mathcal{W}}{\mathcal{V}}\delta_{\Delta}\mathcal{V}+\frac{1}{8\mathcal{V}^{2}}\int\delta\Delta\,\ee^{9\Delta}(9\Theta-12Y\wedge F+6\cos\zeta\,\star f\wedge K).
\end{equation}
Using the torsion and flux decompositions as above, together with $\vol_{8}=\tfrac{1}{7}K\wedge\phi\wedge\psi$, this can be written as 
\begin{equation}
	\delta_{\Delta}\mathcal{W}=\frac{3}{56\mathcal{V}^{2}}\int\delta\Delta\,\ee^{9\Delta}K\wedge\phi\wedge\mathcal{E},
\end{equation}
where the four-form $\mathcal{E}$ is given by
\begin{equation}
	\mathcal{E}=(12\partial_{K}\zeta-21\alpha_{\rep 1}\cos\zeta+2f_{0}\cos\zeta-48\mathcal{V}\mathcal{W})\psi-28\star_{7}\pi_{\rep 1}(F_{3})+28\sin\zeta\,\pi_{\rep 1}(F_{4}).
\end{equation}
This variation vanishes for all $\delta\Delta$ provided $\mathcal{E}=0$.

To isolate the new condition that this variation gives, let the four-forms appearing in \eqref{eq:singlet_1} and \eqref{eq:singlet_2} be $\mathcal E_1$ and $\mathcal E_2$ respectively. The four-form $\mathcal{E}$ can then be expressed as
\begin{equation}
	\mathcal{E}=\frac{4}{3}\mathcal{E}_{1}-\mathcal{E}_{2}+\frac{2}{3}\mathcal{E}_{\Delta},
\end{equation}
where the four-form $\mathcal{E}_{\Delta}$ is
\begin{equation}
	\mathcal{E}_{\Delta}=4\partial_{K}\zeta\,\psi-7\alpha_{\rep 1}\cos\zeta\,\psi-14\star_{7}\pi_{\rep 1}(F_{3})+14\sin\zeta\,\pi_{\rep 1}(F_{4})+2f_{0}\cos\zeta\,\psi-8\mathcal{W}\mathcal{V}\,\psi.\label{eq:delta_Delta}
\end{equation}
Since a critical point should already satisfy $\mathcal{E}_{1}=\mathcal{E}_{2}=0$, the independent constraint coming from the warp-factor variation is simply $\mathcal{E}_{\Delta}=0$.

This rewriting also gives a useful interpretation of this condition. Define the local top-form density
\begin{equation}
W=\tfrac18\left(\Theta-2Y\wedge F+2\cos\zeta\,\star f\wedge K\right),
\qquad\mathcal W=\mathcal V^{-2}\int\ee^{9\Delta}W.
\end{equation}
Using the torsion class and flux decompositions, one finds
\begin{equation}
W-\mathcal W\mathcal V\vol_8=\tfrac1{56}K\wedge\phi\wedge\mathcal E_\Delta.
\end{equation}
Thus, after imposing the two singlet equations, the warp-factor equation is equivalent to
\begin{equation}
\frac{W}{\vol_8}=\mathcal W\mathcal V
=\frac1{\mathcal V}\int\ee^{9\Delta}W.
\end{equation}
Thus the independent part of the warp-factor variation sets the local superpotential density equal to its warped average over the internal manifold.

\subsubsection{$\langle\mathcal{W}\rangle=-m/\mathcal{V}$}

Finally, at a supersymmetric critical point the superpotential itself must satisfy
\begin{equation}
\langle\mathcal{W}\rangle=-\frac{m}{\mathcal{V}},
\end{equation}
as in \eqref{eq:rescaled}. This does not impose any new condition on the internal fields beyond those already obtained from $\langle\delta\mathcal{W}\rangle=0$. Instead, it fixes the inverse AdS scale through 
\begin{equation}
m=-\mathcal{V}\langle\mathcal{W}\rangle.
\end{equation}

\subsection{Recovering the supersymmetry conditions}

We now combine the conditions obtained from the critical-point equations and solve them representation by representation under the local $\Gx 2$ structure. In this way, we recover the full set of supersymmetry conditions.

\subsubsection{$\protect\rep{27}$ conditions}

The conditions on the two $\rep{27}$ torsion classes come from $\delta_{\rep{27}}\phi$ and $\dd Y$, given by \eqref{eq:27_1} and \eqref{eq:27_2} respectively. Solving these for $\star_{7}\delta_{\rep{27}}$ and $\star_{7}\tilde{\delta}_{\rep{27}}$ gives
\begin{equation}
\begin{split}\star_{7}\delta_{\rep{27}} & =\frac{1}{\cos\zeta}\star_{7}\pi_{\rep{27}}(F_{3})+\tan\zeta\,\pi_{\rep{27}}(F_{4}),\\
\star_{7}\tilde{\delta}_{\rep{27}} & =-\frac{1}{\cos\zeta}\pi_{\rep{27}}(F_{4})-\tan\zeta\star_{7}\pi_{\rep{27}}(F_{3}).
\end{split}
\end{equation}
Note that the $\rep{27}$ components of $F_{3}$ and $F_{4}$ remain unconstrained.

\subsubsection{$\protect\rep{14}$ conditions}

The $\rep{14}$ conditions come from $\dd K$ and $\dd Y$, that is, from \eqref{eq:K_tor} and \eqref{eq:14_Y}. Together they simply imply 
\begin{equation}
\gamma_{\rep{14}}=\eta_{\rep{14}}=0.
\end{equation}

\subsubsection{$\protect\rep 7$ conditions}

The $\rep 7$ conditions arise from $\dd K$, $\dd Y$, $\delta_{\rep 7}K$ and the two $\delta_{\rep 7}\phi$ variations. These are given by Equations \eqref{eq:K_tor}, \eqref{eq:beta7}, \eqref{eq:delta_K_7}, \eqref{eq:delta_phi_7} and \eqref{eq:delta_phi_7_2} respectively. Solving these gives all of the $\rep 7$ torsion classes, the two $\rep 7$ components of $F$, and the $\rep 7$ component of $f$:
\begin{equation}
\begin{aligned}\tilde{\beta}_{\rep 7} & =3\dd_{7}\Delta-\tan\zeta\,\dd_{7}\zeta, & \star_{7}\pi_{\rep 7}(F_{3}) & =\tfrac{1}{2}\ee^{-3\Delta}\dd_{7}(\ee^{3\Delta}\cos\zeta)\wedge\phi,\\
\gamma_{\rep 7} & =0, & \pi_{\rep 7}(F_{4}) & =\tfrac{1}{2}\phi\wedge\dd_{7}\zeta,\\
\beta_{\rep 7} & =-\tfrac{3}{2}\dd_{7}\Delta, & f_{1} & =\ee^{-3\Delta}\dd_{7}(\ee^{3\Delta}\sin\zeta),\\
\star_{7}\tilde{\delta}_{\rep 7} & =\frac{1}{\cos\zeta}\dd_{7}\zeta\wedge\phi-\tfrac{1}{2}\ee^{-3\Delta}\dd_{7}(\ee^{3\Delta}\sin\zeta)\wedge\phi.
\end{aligned}
\end{equation}

\subsubsection{$\protect\rep 1$ conditions}

The singlet conditions come from $\dd Y$, $\delta\zeta$, $\delta_{\rep 1}K$, $\delta_{\rep 1}\phi$ and $\delta\Delta$, and are given by \eqref{eq:14_Y}, \eqref{eq:xi_1}, \eqref{eq:singlet_1}, \eqref{eq:singlet_2} and \eqref{eq:delta_Delta}.  Together with the identification $m=-\mathcal{W}\mathcal{V}$, these determine $f_{0}$ in terms of $\zeta$, $\Delta$ and $m$, and hence fix the remaining singlet torsion classes and singlet flux components: 
\begin{equation}
\begin{split}\alpha_{\rep 1} & =\frac{2}{7}\cos\zeta\left(\partial_{K}\zeta+12m-3\tan\zeta\,\partial_{K}\Delta+2(\partial_{K}\zeta+8m)\tan^{2}\zeta\right),\\
\tilde{\alpha}_{\rep 1} & =\frac{1}{7}\left(-12\partial_{K}\Delta+(\partial_{K}\zeta+8m)\tan\zeta\right),\\
\pi_{\rep 1}(F_{3}) & =\frac{2}{7}\left(\partial_{K}\zeta-2m\right)\phi,\\
\pi_{\rep 1}(F_{4}) & =\frac{2}{7}\left(4m\sin\zeta-\ee^{-3\Delta}\partial_{K}(\ee^{3\Delta}\cos\zeta)\right)\psi,\\
f_{0} & =\ee^{-3\Delta}\partial_{K}(\ee^{3\Delta}\sin\zeta)+4m\cos\zeta.
\end{split}
\end{equation}

\subsubsection{Equivalence with eleven-dimensional supersymmetry}

We can now state the main result of this section. The critical-point conditions of the off-shell superpotential determine the intrinsic torsion of the local $\Gx 2$ structure completely: the exterior derivatives $\dd K$, $\dd\phi$ and $\dd\psi$ are fixed in terms of the fluxes $F$ and $f$, the warp factor $\Delta$, and the function $\zeta$. In addition, all components of the flux are fixed except for the two $\rep{27}$ components of $F$, which remain unconstrained by supersymmetry.

Comparing the results obtained above with the analysis of \cite{Martelli:2003ki}, reviewed in Section \ref{subsec:Martelli_review}, we find that the critical-point equations reproduce precisely the system \eqref{eq:sufficient_1}--\eqref{eq:sufficient_6}. That system is known to be necessary and sufficient for a solution to the Killing spinor equations \eqref{eq:KSE-1} and \eqref{eq:KSE-2}. Thus the conditions 
\begin{equation}
\langle\delta\mathcal{W}\rangle=0,\qquad\langle\mathcal{W}\rangle=-\frac{m}{\mathcal{V}},
\end{equation}
are equivalent to the full set of supersymmetry conditions on the eleven-dimensional background.

The Bianchi identities \eqref{eq:bianchi1} are automatically satisfied when the fluxes are expressed using local potentials $A$ and $\tilde{A}$. When using the fluxes $(F,f)$, they must be imposed separately. As reviewed in Section \ref{subsec:Eleven-dimensional-supergravity-}, the Killing spinor equations plus \eqref{eq:bianchi1} then imply the remaining two-derivative bosonic equations. Within a given flux sector, we then have the equivalence
\[
\begin{array}{c}
\begin{array}{c}
\langle\delta\mathcal{W}\rangle=0,\quad\langle\mathcal{W}\rangle=-m\mathcal{V}^{-1},\\
\dd(\ee^{3\Delta}F)=0,\quad\ee^{-6\Delta}\dd(\ee^{6\Delta}\star f)+\tfrac{1}{2}F\wedge F=0
\end{array}\\
\Updownarrow\\
(\Delta,g,F,f)\text{ defines a supersymmetric eleven-dimensional background.}
\end{array}
\]
More precisely, this should be understood locally in terms of the existence of a nowhere-vanishing spinor defining the local $\Gx 2$ structure together with the differential conditions above. Any additional global constraints, such as flux quantisation or tadpole conditions, must be imposed separately.

\section{The global \texorpdfstring{$\Spin7$}{Spin(7)} limit\label{sec:The-global-}}

We saw in Section \ref{sec:The--superpotential} that the general off-shell superpotential reduces on the $\Spin 7$ locus to 
\begin{equation}
	\left.\mathcal{W}\right|_{\Spin 7}=-\frac{1}{4\mathcal{V}^{2}}\int\ee^{9\Delta}Y\wedge F,
\end{equation}
where $Y$ is the self-dual Cayley form defined by $\xi_+$. At first sight, one might expect that varying this simpler functional should reproduce the supersymmetry conditions for M-theory compactified on an eight-manifold with a global $\Spin 7$ structure. As we now explain, this is not quite correct.

Suppose one varies the above functional directly. The admissible deformations are then those that move one within the space of $\Spin 7$ structures, together with variations of the three-form potential and warp factor. Concretely, one should consider the variations $(\delta_{\rep 1}Y,\delta_{\rep 7}Y,\delta_{\rep{35}}Y,\delta A,\delta\Delta)$, where $\delta_{\rep{27}}Y$ is not allowed because it does not lie in the tangent space to the space of $\Spin 7$ structures.\footnote{Equivalently, the tangent space of admissible $\Spin 7$ forms is spanned by $\rep 1+\rep 7+\rep{35}$. See also~\cite{1404.2460}.} Varying $\left.\mathcal{W}\right|_{\Spin 7}$ with respect to these, one finds\footnote{Here the subscripts denote $\Spin 7$ representations. Recall that a generic four-form on a manifold with a $\Spin 7$ structure decomposes as $\rep 1_{+}+\rep 7_{+}+\rep{27}_{+}+\rep{35}_{-}$, where $+$ or $-$ denote self-duality or anti-self-duality with respect to the Hodge star.}
\begin{gather}
\pi_{\rep1}(F)=-\tfrac87\mathcal W\mathcal V\,Y,\qquad
\pi_{\rep7}(F)=0,\qquad\pi_{\rep{35}}(F)=0,\\
\dd(\ee^{6\Delta}Y)=0,\qquad
Y\wedge F+12\,\mathcal W\mathcal V\vol_8=0,
\end{gather}
where we again identify $m=-\mathcal{V}\langle\mathcal{W}\rangle$. Substituting the first expression into the last and using $Y\wedge Y=14\vol_{8}$ then gives $m=0$ and hence $\pi_{\rep 1}(F)=0$.
Thus, critical points of the restricted functional reproduce only
\begin{equation}
	\dd(\ee^{6\Delta}Y)=0,\qquad\pi_{\rep 1}(F)=\pi_{\rep 7}(F)=\pi_{\rep{35}}(F)=0,\qquad m=0.
\end{equation}
This is not the full set of supersymmetry conditions for the warped $\Spin 7$ case. In particular, supersymmetry allows a non-zero electric flux $f=3\,\dd\Delta$ as shown in \cite{Acharya:2002vs,Martelli:2003ki}, and this information is absent from \eqref{eq:spin7_superpotential}. The reason is that \eqref{eq:spin7_superpotential} is obtained only after restricting the general off-shell functional to the $\Spin 7$ locus. If one first restricts the functional and only then varies it, one loses the off-shell directions needed to recover the full set of supersymmetry conditions.

The missing equation can be obtained by allowing variations transverse to the $\Spin7$ locus. At $\sin\zeta=1$, one keeps the metric, warp factor and potentials fixed and varies the negative-chirality component of the unit spinor. The induced one-form $k=\delta\bar K$ is then arbitrary with the various spinor bilinears varying as
\begin{equation}
\delta(\sin\zeta)=0,\qquad\delta Y=0,\qquad
\delta\bar\phi=-\imath_{k^\sharp}Y.
\end{equation}
Varying the superpotential then gives
\begin{equation}
\delta_\perp\mathcal W=-\frac1{8\mathcal V^2}\int\ee^{9\Delta}
\bigl[-4\dd\star k+(\imath_{k^\sharp}Y)\wedge\dd Y-2\star f\wedge k\bigr].
\end{equation}
Recall that the tangential $\Spin7$ variations already give $\dd Y=-6\,\dd\Delta\wedge Y$. Using
$(\imath_{k^\sharp}Y)\wedge Y=7\imath_{k^\sharp}\vol_8=-7\star k$, integration by parts yields
\begin{equation}
\delta_\perp\mathcal W=\frac1{4\mathcal V^2}\int\ee^{9\Delta}
\star(f-3\dd\Delta)\wedge k.\label{eq:spin7_transverse_variation}
\end{equation}
Requiring this to vanish for arbitrary $k$ then imposes $f=3\,\dd\Delta$, recovering the condition lost by restricting the functional before varying it.

The global $\Spin 7$ structure considered above is defined by the positive-chirality spinor $\xi_+$. Thus this limit corresponds to setting $\xi_-=0$, or equivalently $\sin\zeta=+1$. The other branch, $\sin\zeta=-1$, corresponds to choosing $\xi_-$ non-vanishing. Specialising the differential and algebraic conditions of Appendix \ref{app:bilinears} to the $\sin\zeta=+1$ branch, the surviving conditions are 
\begin{equation}
	\begin{gathered}
		f=3\,\dd\Delta,\qquad0=-\star F+F-4mY,\qquad\ee^{-6\Delta}\dd(\ee^{6\Delta}Y)=0,\\
		0=2m+\tfrac{1}{6}Y\lrcorner F,\qquad0=\tfrac{1}{36}Y_{a}{}^{bcd}F_{bcde}e^{ae},
	\end{gathered}
\end{equation}
which together imply\footnote{Choosing the other branch, $\sin\zeta=-1$, sends $f\to -f$ and flips the sign of the Hodge-star eigenvalue for both the Cayley form $Y$ and the allowed component of the four-form flux $F$. Equivalently, it gives the same $\Spin 7$ supersymmetry conditions written with the opposite orientation on $X$.}
\begin{equation}
	f=3\,\dd\Delta,\qquad\dd(\ee^{6\Delta}Y)=0,\qquad\pi_{\rep 1}(F)=\pi_{\rep 7}(F)=\pi_{\rep{35}}(F)=0,\qquad m=0.
\end{equation}
These are the standard two-derivative $\Spin7$ supersymmetry conditions: the external space is Minkowski, the internal metric is conformal to a torsion-free $\Spin7$ metric, and the magnetic flux lies in the $\rep{27}$~\cite{Acharya:2002vs,Martelli:2003ki,Becker:2000jc}. As discussed in Section \ref{subsec:Eleven-dimensional-supergravity-}, the existence of a compact flux vacuum requires additional global conditions.

\section{Discussion\label{sec:Discussion}}

We have derived the real superpotential functional $\mathcal{W}$ governing warped compactifications of eleven-dimensional supergravity on eight-manifolds to Minkowski or AdS vacua preserving $N=1$ supersymmetry in three dimensions. We obtained this superpotential in two independent ways: first from the dimensional reduction of the gravitino variation and second from the reduction of the eleven-dimensional gravitino action. We then showed that the critical points of $\mathcal{W}$ reproduce the full set of supersymmetry conditions implied by the eleven-dimensional Killing spinor equations. We also commented on the $\Spin 7$ limit: although the general superpotential reduces to the known functional for $\Spin 7$ manifolds, recovering the full set of supersymmetry conditions requires varying the general off-shell functional before restricting to the $\Spin 7$ locus.

A natural question is what happens when one includes quantum corrections. For compact $X$, the supersymmetry equations must also be combined with an integrated charge constraint that leads to the no-go described in Section~\ref{subsec:Eleven-dimensional-supergravity-}. We plan to address the compact case in future work. Furthermore, non-perturbative corrections to lower-dimensional superpotentials often admit a higher-dimensional interpretation. For example, in type IIB warped Calabi--Yau compactifications, Euclidean D3-brane effects deform the conventional complex structure into a generalised complex structure~\cite{0707.1038}. It would be very interesting to understand whether quantum corrections can be incorporated into the real superpotential derived here, and whether they admit a direct interpretation in eleven dimensions. Such corrections are likely to be important for any attempt to stabilise moduli in these backgrounds. 

An instructive comparison is the work of Becker, Robbins and Witten~\cite{1404.2460}, who showed that compactifications on exceptional holonomy manifolds can be deformed to maintain supersymmetry in the presence of corrections. For M-theory, their argument applies to all finite orders in the inverse-radius expansion, but does not exclude non-perturbative supersymmetry breaking. Whether an analogous statement holds for the more general warped flux backgrounds considered here is an interesting question.

A second question concerns the moduli and masses of supersymmetric solutions. Infinitesimal supersymmetric deformations lie in the kernel of the linearised critical-point equations, subject to the flux constraints and modulo gauge equivalence~\cite{delaOssa:2015maa}. Identifying genuine moduli beyond linear order also requires analysing higher-order obstructions, while computing physical masses requires the field-space metric $G_{IJ}$ obtained from the eleven-dimensional kinetic terms. Even without computing this metric explicitly, the real superpotential implies the scalar BF bound. In the conventions of \eqref{eq:3d_Action}, define
\[
\mathcal W_0=\langle\mathcal W\rangle,\qquad
H_{IJ}=\langle\nabla_I\partial_J\mathcal W\rangle.
\]
At a supersymmetric critical point, one has
\begin{equation}
	\langle\nabla_I\partial_J V\rangle
	=2H_{IK}G^{KL}H_{LJ}-8\mathcal W_0H_{IJ}.
\end{equation}
Because the scalar kinetic term is $-G_{IJ}\partial\phi^I\partial\phi^J$, the physical mass operator is
\begin{equation}
	(M^2)^I{}_J
	=(\mathsf H^2)^I{}_J-4\mathcal W_0\mathsf H^I{}_J,
	\qquad
	\mathsf H^I{}_J=G^{IK}H_{KJ}.
\end{equation}
Given a positive-definite metric on the physical scalar modes, $\mathsf H$ is self-adjoint, so an eigenvalue $\lambda$ gives
\begin{equation}
	M^2=(\lambda-2\mathcal W_0)^2-4\mathcal W_0^2
	\geq-\frac{1}{L_{\mathrm{AdS}}^2},
\end{equation}
where $L_{\mathrm{AdS}}^{-2}=4\mathcal W_0^2$. Thus the scalar spectrum automatically satisfies the $\AdS3$ BF bound. This does not, however, determine the spectrum itself: calculating the actual masses and their multiplicities still requires the field-space metric. In particular, a massless mode can have either $\lambda=0$ or $\lambda=4\mathcal W_0$, but only the former is an infinitesimal supersymmetric modulus, and it may still be obstructed at higher order.

More broadly, finding the superpotential without first performing a Kaluza--Klein truncation is a necessary step toward constructing a finite-dimensional effective theory for the light modes of the theory. A next step would be to identify a consistent truncation of the infinite set of fluctuations to a finite set of fields. In type II compactifications on $\SU 3\times\SU 3$ structure manifolds, this is aided by the special geometric structures that appear on the space of fields~\cite{Grana:2005ny,Grana:2006hr,Ashmore:2015joa,Ashmore:2016qvs,Ashmore:2019qii}. For the geometries studied here, the situation is more subtle: the background is described in terms of a local $\Gx 2$ structure which may enhance to $\Spin 7$ on subsets of $X$, and it is not clear that the corresponding space of structures admits a comparably useful description. This suggests that a more global reformulation of the geometry may be needed, perhaps in terms of the generalised $\Spin 7\times\Spin 7$ structures discussed in~\cite{hep-th/0511047}. 

The variational structure also suggests numerical approaches to constructing these backgrounds. Since supersymmetric vacua sit at critical points of $\mathcal{W}$, a gradient flow on the space of local $\Gx2$ structures and fluxes may converge to a solution from generic initial data, in the spirit of Laplacian flow for torsion-free $\Gx2$ structures~\cite{lotay2019stability} or the anomaly-flow construction of heterotic backgrounds~\cite{Phong:2016ggz}. Unfortunately, convergence cannot be inferred from the existence of the functional alone: supersymmetric critical points may be saddles, and gauge directions and flux constraints must be accounted for. Regardless, it would be worth understanding whether the resulting flow is a previously studied geometric flow on the space of $\Gx2$ structures, or a genuinely new one fixed by the flux terms in $\mathcal{W}$.


The AdS$_3$ vacua described here are dual to two-dimensional $N=(1,0)$ SCFTs, with the central charge fixed by $\langle\mathcal{W}\rangle=-m/\mathcal{V}$. For $N=(0,2)$ backgrounds, $c$-extremisation~\cite{Benini:2012cz,Benini:2013cda} fixes the central charge by extremising over a trial $\Uni1_\text{R}$ symmetry; its geometric counterpart, GK geometry~\cite{Couzens:2018wnk}, reproduces this extremisation over a finite-dimensional family of geometric data, without solving the full supersymmetry conditions. The minimal $N=(1,0)$ algebra, however, contains no $\Uni1_\text{R}$ current, so there is no trial R-symmetry to extremise over. A more modest question remains, however: whether restricting $\mathcal{W}$ to a finite-dimensional family of trial data fixes $c$ without needing the full supersymmetry conditions. In particular, one would want to know what geometric data on $X$ plays the role that the Reeb vector plays in GK geometry.

An intriguing possible application of the more general local \(G_2\) structures studied here is to revisit Witten's proposal for obtaining a nonsupersymmetric four-dimensional vacuum from a supersymmetric three-dimensional theory~\cite{hep-th/9409111,hep-th/9506101}, as suggested in the context of F-theory by Vafa~\cite{hep-th/9602022}. The additional freedom provided by nonchiral internal spinors, intrinsic torsion and fluxes may permit degenerating torus fibrations with nontrivial $\SL{2,\mathbb Z}$ monodromies, whose F-theory limits need not preserve four-dimensional supersymmetry. The real superpotential derived here provides a starting point for investigating families of three-dimensional \(N=1\) Minkowski vacua approaching such a limit. A crucial question is whether the additional states becoming light as the elliptic fiber collapses restore supersymmetry, as happens in certain previously studied Spin(7) constructions in the weak-coupling, infinite-interval limit~\cite{1309.2287}, or whether more general backgrounds can realize this scenario.

Finally, it remains to understand how this story changes for compactifications preserving $N=2$ supersymmetry in three dimensions. With more supersymmetry, one expects a more rigid off-shell geometry, closer in spirit to four-dimensional $N=1$ compactifications, and it is conceivable that this additional structure could impose stronger constraints on perturbative corrections to the effective theory. We leave this question for future investigation.

\subsection*{Acknowledgements}

We thank Xenia de la Ossa, Charles Strickland-Constable, Eirik Eik Svanes, David Tennyson, Daniel Waldram and Jock McOrist for interesting and insightful conversations. SS is supported in part by NSF Grant No.~PHY2412985. The authors used OpenAI ChatGPT and Codex to assist with manuscript editing and to check selected calculations and derivations. All AI-assisted output was reviewed and independently verified by the authors.

\appendix

\section{Conventions\label{app:conventions}}

\subsection{Eleven-dimensional supergravity\label{subsec:Eleven-dimensional-supergravity}}

We follow the conventions of \cite{Martelli:2003ki}. These agree with \cite{Gauntlett:2002fz,Coimbra:2012af} after sending $\Gamma_{M}\to-\Gamma_{M}$.\footnote{This is consistent with the observation that \cite{Gauntlett:2002fz,Coimbra:2012af} take $\Gamma_{11}=+\id_{32}$. In \cite{Martelli:2003ki} and in our conventions, the decomposition of the eleven-dimensional gamma matrices, given in \eqref{eq:gamma_decomposition}, instead implies $\Gamma_{11}=-\id_{32}$.} The equations of motion of eleven-dimensional supergravity, to quadratic order in fermions, are
\begin{align}
\hat{R}_{MN}-\tfrac{1}{12}\left(G_{MPQR}G_{N}{}^{PQR}-\tfrac{1}{12}\hat{g}_{MN}G_{PQRS}G^{PQRS}\right) & =0,\label{eq:eq1}\\
\dd\mathop{\hat{\star}}G+\tfrac{1}{2}G\wedge G & =0,\\
\Gamma^{MNP}\hat{\nabla}_{N}\psi_{P}-\tfrac{1}{96}\left(\Gamma^{MNPQRS}G_{PQRS}+12G^{MN}{}_{PQ}\Gamma^{PQ}\right)\psi_{N} & =0.\label{eq:eq3}
\end{align}
with the usual Bianchi identity $\dd G=0$. As usual, the equations of motion receive higher order corrections. For example, the equation of motion for $G$ picks up a contribution $X_{8}$ on the right-hand side, given by
\begin{equation}
\dd\mathop{\hat{\star}}G+\tfrac{1}{2}G\wedge G=-\beta X_{8},\label{eq:flux_X8}
\end{equation}
with 
\begin{equation}
X_{8}=\frac{1}{192}(p^{2}_{1}-4p_{2}),\qquad\beta=\frac{2\pi}{T_{5}},\qquad T_{5}=\frac{(2\pi)^{1/3}}{(2\kappa^{2})^{2/3}},\label{eq:X8_definition}
\end{equation}
where $p_{i}$ is the $i^{\text{th}}$ Pontryagin form and the M2-brane tension is set to one.

The minimal spinor in eleven dimensions is Majorana. Throughout this work, we take $\eta$ to be a Majorana spinor of $\Spin{1,10}$ and let $\Gamma_{M}$ furnish a representation of $\op{Cliff}(1,10)$:
\begin{equation}
\{\Gamma_{M},\Gamma_{N}\}=2\hat{g}_{MN}.
\end{equation}
We take the spacetime signature to be $(-,+,\dots,+)$ so that $\Gamma_{M}$ is hermitian for $M\neq0$ and anti-hermitian for $M=0$. The Dirac adjoint of a spinor is defined as
\begin{equation}
\bar{\eta}=\eta^{\dagger}\Gamma^{0}\equiv\eta^{\transpose}C,
\end{equation}
where the second equality is true for Majorana spinors, and $C$ is the charge conjugation matrix $C=-C^{\transpose}=\Gamma^{0}$.

The bosonic and fermionic contributions to the action that lead to the equations of motion in \eqref{eq:eq1}--\eqref{eq:eq3} are
\begin{align}
S_{\text{B}} & =\frac{1}{2\kappa^{2}}\int\left(\hat{\star}\hat{R}-\tfrac{1}{2}G\wedge\hat{\star}G-\tfrac{1}{6}C\wedge G\wedge G\right),\label{eq:11d_bosonic_action}\\
\begin{split}S_{\text{F}} & =\frac{1}{\kappa^{2}}\int\sqrt{-\hat{g}}\biggl(-\bar{\psi}_{M}\Gamma^{MNP}\hat{\nabla}_{N}\psi_{P}+\tfrac{1}{96}G_{PQRS}\bar{\psi}_{M}\Gamma^{MPQRSN}\psi_{N}\\
 & \phantom{=\frac{1}{\kappa^{2}}\int\sqrt{-\hat{g}}\biggl(}+\tfrac{1}{8}G_{MNPQ}\bar{\psi}^{M}\Gamma^{NP}\psi^{Q}\biggr).
\end{split}
\label{eq:11d_fermionic_action}
\end{align}
The total action is supersymmetric, to quadratic order in fermions, under the supersymmetry variations
\begin{align}
\delta\hat{g}_{MN} & =-2\bar{\eta}\Gamma_{(M}\psi_{N)},\\
\delta C_{MNP} & =3\bar{\eta}\Gamma_{[MN}\psi_{P]},\\
\delta\psi_{M} & =\hat{\nabla}_{M}\eta-\tfrac{1}{288}(\Gamma_{M}{}^{PQRS}-8\delta^{P}_{M}\Gamma^{QRS})G_{PQRS}\eta,
\end{align}
where $\eta$ is a Majorana supersymmetry parameter.

\subsection{Geometry\label{subsec:Geometry}}

We define the Hodge star on the eight-manifold $X$ by
\begin{equation}
\alpha\lrcorner\beta\vol_{8}=\star\alpha\wedge\beta,
\end{equation}
where $\alpha$ and $\beta$ are both $p$-forms. On an eight-manifold, the Hodge star squares to give $\star^{2}\alpha=(-1)^{p}\alpha$.

We will also need some identities involving the divergence of a tensor. For a vector field $v$, we have 
\begin{equation}
\mathcal{L}_{v}\text{vol}_{8}=(\op{div}v)\vol_{8}\equiv-\dd\star v^{\flat},
\end{equation}
where $\flat$ denotes lowering the index of $v$ with the metric.\footnote{We often drop this notation and leave the lowering implicit where it will not cause confusion.} We then have $\op{div}v\equiv\nabla\lrcorner v^{\flat}=-\vol^{-1}_{8}\dd\star v^{\flat}$, where $\nabla$ is the Levi-Civita connection associated to the metric on $X$ and $\lrcorner$ denotes the interior product.\footnote{The interior product is often denoted by $\imath$, so that this identity would read $\imath_{\nabla}v^{\flat}=-\vol^{-1}_{8}\dd\star v^{\flat}$.} Extending this to a $p$-form $\alpha$ gives
\begin{equation}
\nabla\lrcorner\alpha=-\star(\nabla\wedge\star\alpha)=-\star\dd\star\alpha.
\end{equation}
Contracting with a $(p-1)$-vector $\beta$ then gives the identity
\begin{equation}
\beta\lrcorner(\nabla\lrcorner\alpha)=-\beta\lrcorner\star\dd\star\alpha=-\vol^{-1}_{8}\beta\wedge\dd\star\alpha.
\end{equation}

\subsection{Clifford algebra and spinors}\label{app:clifford}

A thorough review of gamma matrices in any dimension can be found in \cite{1905.00429}. Some useful commutator and anticommutator identities can be found in \cite{0711.1436}. The relevant Clifford algebra in eleven dimensions is $\op{Cliff}(1,10)$, which we take to be generated by $\Gamma_{M}$ such that
\begin{equation}
\{\Gamma_{M},\Gamma_{N}\}=2\hat{g}_{MN},\qquad\Gamma_{11}=\Gamma_{0}\dots\Gamma_{10}=-\id_{32}.
\end{equation}

\subsubsection{Three dimensions}

An explicit representation of the three-dimensional Clifford algebra is given by the Pauli matrices:
\begin{equation}
\gamma_{0}=\ii\sigma_{1},\qquad\gamma_{1}=\sigma_{2},\qquad\gamma_{2}=\sigma_{3}.
\end{equation}
Here $\gamma_{0}$ is anti-hermitian while $\gamma_{1}$ and $\gamma_{2}$ are hermitian, consistent with
\begin{equation}
\{\gamma_{\mu},\gamma_{\nu}\}=2\eta_{\mu\nu},\qquad\eta_{\mu\nu}=\op{diag}(-1,1,1).
\end{equation}
One can check that\footnote{This can be derived using $\gamma_{\mu}=(\ii\sigma_{1},\sigma_{2},\sigma_{3})$ and $\eta_{\mu\nu}=\op{diag}(-1,1,1)$, in agreement with the conventions of \cite{Martelli:2003ki}.}
\begin{equation}
\gamma_{\mu}=\tfrac{1}{2}\epsilon_{\mu\nu\rho}\gamma^{\nu\rho},\qquad\gamma_{\mu\nu}=-\epsilon_{\mu\nu\rho}\gamma^{\rho},
\end{equation}
where the Levi-Civita symbol in $(-1,1,1)$ signature obeys $\epsilon_{012}=-\epsilon^{012}$.

\subsubsection{Eight dimensions}

We choose a real symmetric representation of the eight-dimensional Clifford algebra on $\mathbb{R}^{16}$, so that the gamma matrices are $16\times16$ real symmetric matrices with $\gamma^{2}_{9}=\id_{16}$ and
\begin{equation}
\{\gamma_{m},\gamma_{n}\}=2\delta_{mn},\qquad\gamma_{9}=\gamma_{1}\dots\gamma_{8}.
\end{equation}
An explicit realisation, useful for calculations, is given by
\begin{align}
\begin{split}\gamma_{1} & =\sigma_{1}\otimes\id_{2}\otimes\id_{2}\otimes\id_{2},\\
\gamma_{3} & =\sigma_{2}\otimes\sigma_{2}\otimes\sigma_{1}\otimes\id_{2},\\
\gamma_{5} & =\sigma_{2}\otimes\sigma_{1}\otimes\id_{2}\otimes\sigma_{2},\\
\gamma_{7} & =\sigma_{2}\otimes\id_{2}\otimes\sigma_{2}\otimes\sigma_{1},
\end{split}
 & \begin{split}\gamma_{2} & =\sigma_{3}\otimes\id_{2}\otimes\id_{2}\otimes\id_{2},\\
\gamma_{4} & =\sigma_{2}\otimes\sigma_{2}\otimes\sigma_{3}\otimes\id_{2},\\
\gamma_{6} & =\sigma_{2}\otimes\sigma_{3}\otimes\id_{2}\otimes\sigma_{2},\\
\gamma_{8} & =\sigma_{2}\otimes\id_{2}\otimes\sigma_{2}\otimes\sigma_{3},
\end{split}
\label{eq:gamma_realisation}
\end{align}
where the $\sigma_{i}$ are the Pauli matrices and $\id_{2}$ is the $2\times2$ identity matrix. With this choice, the highest-rank gamma matrix is 
\[
\gamma_{9}=\sigma_{2}\otimes\sigma_{2}\otimes\sigma_{2}\otimes\sigma_{2}.
\]
In this realisation, the intertwiners can be taken to be trivial.\footnote{This is the case $\eta=\epsilon=-1$ in the notation of \cite{1905.00429}.} In particular, $\xi^{\charge}=\xi^{*}$ and $\bar{\xi}=\xi^{\dagger}$ for a spinor $\xi$ in eight dimensions.

\subsubsection{$11\to3+8$ decomposition\label{subsec:Metric-and-gamma}}

Given our ansatz for the eleven-dimensional metric in \eqref{eq:11_metric}, the eleven-dimensional gamma matrices decompose as
\begin{align}
\begin{split}\Gamma^{\mu} & =\ee^{-\Delta}(\gamma^{\mu}\otimes\gamma_{9}),\\
\Gamma_{\mu} & =\ee^{\Delta}(\gamma_{\mu}\otimes\gamma_{9}),
\end{split}
 & \begin{split}\Gamma^{m} & =\ee^{-\Delta}(\id_{2}\otimes\gamma^{m}),\\
\Gamma_{m} & =\ee^{\Delta}(\id_{2}\otimes\gamma_{m}).
\end{split}
\label{eq:gamma_decomposition}
\end{align}

\subsubsection{Spinors}

We take $\xi_{\pm}$ to be real chiral spinors in eight dimensions, while $\psi$ is Majorana in three dimensions:
\begin{equation}
\gamma_{9}\xi_{\pm}=\pm\xi_{\pm},\qquad\psi^{*}=\gamma_{2}\psi.
\end{equation}
These conditions ensure that $\eta$ is Majorana in eleven dimensions. Note that $\psi$ has two real components and $\xi_{\pm}$ are sixteen component spinors, each with eight real degrees of freedom.

\subsubsection{Gamma matrix identities}

We collect some useful properties of gamma matrices in eight dimensions. First, we have the (anti)commutation identity
\begin{equation}
\gamma_{9}\gamma_{(n)}=\begin{cases}
\gamma_{(n)}\gamma_{9} & n=\text{even},\\
-\gamma_{(n)}\gamma_{9} & n=\text{odd},
\end{cases}
\end{equation}
for a rank-$n$ gamma matrix, so that, for example, $\gamma_{9}$ anticommutes with rank-one gamma matrices. As matrices, the rank-$n$ gamma matrices are
\begin{equation}
\gamma_{(n)}=\begin{cases}
\text{symmetric} & n=0,1,4,5,8,\\
\text{antisymmetric} & n=2,3,6,7.
\end{cases}
\end{equation}
This symmetry table constrains which bilinears are non-vanishing and their reality properties. For example, for any spinor $\xi$,
\begin{equation}
\bar{\xi}\gamma_{(n)}\xi=\begin{cases}
\text{real} & (-1)^{n(n-1)/2}=1,\\
\text{imaginary} & (-1)^{n(n-1)/2}=-1.
\end{cases}
\end{equation}

For Weyl spinors $\xi_{1,2}$, spinors of opposite chirality are orthogonal, while odd-rank gamma matrices flip chirality. Hence
\begin{equation}
\bar{\xi}_{2}\gamma_{(n)}\xi_{1}=0\quad\text{if }\quad\begin{cases}
\text{\ensuremath{\xi_{1}} and \ensuremath{\xi_{2}} have same chirality and \ensuremath{n} is odd,}\\
\text{\ensuremath{\xi_{1}} and \ensuremath{\xi_{2}} have opposite chirality and \ensuremath{n} is even.}
\end{cases}
\end{equation}
For Majorana spinors $\xi_{1,2}$, one instead has
\begin{equation}\label{eq:majorana_identities}
\begin{split}\bar{\xi}_{1}\gamma_{(n)}\xi_{2} & =\begin{cases}
+\bar{\xi}_{2}\gamma_{(n)}\xi_{1} & n=0,1,4,5,8,\\
-\bar{\xi}_{2}\gamma_{(n)}\xi_{1} & n=2,3,6,7,
\end{cases}\\
\bar{\xi}_{1}\gamma_{9}\gamma_{(n)}\xi_{2} & =\begin{cases}
+\bar{\xi}_{2}\gamma_{9}\gamma_{(n)}\xi_{1} & n=0,3,4,7,8,\\
-\bar{\xi}_{2}\gamma_{9}\gamma_{(n)}\xi_{1} & n=1,2,5,6.
\end{cases}
\end{split}
\end{equation}
When computing the superpotential, the following Fierz identity is useful:
\begin{equation}
\begin{split}2^{4}\,\xi\chi^{\dagger} & =\id_{16}\chi^{\dagger}\xi+\gamma_{m}\chi^{\dagger}\gamma^{m}\xi-\tfrac{1}{2!}\gamma_{mn}\chi^{\dagger}\gamma^{mn}\xi-\tfrac{1}{3!}\gamma_{mnp}\chi^{\dagger}\gamma^{mnp}\xi+\tfrac{1}{4!}\gamma_{mnpq}\chi^{\dagger}\gamma^{mnpq}\xi\\
 & \eqspace+\gamma_{9}\chi^{\dagger}\gamma_{9}\xi-\gamma_{m}\gamma_{9}\chi^{\dagger}\gamma^{m}\gamma_{9}\xi-\tfrac{1}{2!}\gamma_{mn}\gamma_{9}\chi^{\dagger}\gamma^{mn}\gamma_{9}\xi+\tfrac{1}{3!}\gamma_{mnp}\gamma_{9}\chi^{\dagger}\gamma^{mnp}\gamma_{9}\xi\\
 & \equiv\sum^{8}_{p=0}\frac{1}{p!}\gamma_{m_{1}\dots m_{p}}\chi^{\dagger}\gamma^{m_{p}\dots m_{1}}\xi.
\end{split}
\label{eq:fierz}
\end{equation}

\section{More on the \texorpdfstring{$\Gx 2$}{G\_2} structure\label{app:G2-bilinears}}

\subsection{Spinor projections}

A $\Gx 2$ structure in eight dimensions may be defined by starting with a pair of spinors and then imposing certain projections, as described in \cite[Appendix B]{Gauntlett:2003cy}. We reproduce these here for completeness and to fix our conventions. Locally, a $\Gx 2$ structure is defined by a pair of positive- and negative-chirality Majorana--Weyl spinors, $\xi_{\pm}$. Given a choice of basis in which the gamma matrices are all real and symmetric, the spinors $\xi_{\pm}$ are real sixteen-component objects. One then imposes the conditions
\begin{equation}
\begin{gathered}\gamma_{9}\xi_{\pm}=\pm\xi_{\pm},\qquad\xi^{\transpose}_{\pm}\xi_{\pm}=1\pm\sin\zeta,\\
\gamma_{1234}\xi_{\pm}=\gamma_{1256}\xi_{\pm}=\gamma_{1357}\xi_{\pm}=-\xi_{\pm},
\end{gathered}
\end{equation}
where the indices refer to an orthonormal frame $\{e^{m}\}$ for the metric on $X$. Although the Weyl spinors $\xi_{\pm}$ can have zeros, the non-chiral combinations
\begin{equation}
\epsilon_{\pm}=\frac{1}{\sqrt{2}}(\xi_{+}\pm\xi_{-}),\qquad\gamma_{9}\epsilon_{\pm}=\epsilon_{\mp},
\end{equation}
are nowhere vanishing, with their norms and inner products given by
\begin{equation}
\epsilon^{\transpose}_{+}\epsilon_{+}=\epsilon^{\transpose}_{-}\epsilon_{-}=1,\qquad\epsilon^{\transpose}_{+}\epsilon_{-}=\epsilon^{\transpose}_{-}\epsilon_{+}=\sin\zeta.
\end{equation}

Given the above projections, the forms defining the associated $\Gx 2$ structure on $X$ are
\begin{align}
K & =-e^{8},\\
\phi & =e^{127}-e^{136}-e^{145}-e^{235}+e^{246}+e^{347}+e^{567}\\
\psi=\star_{7}\phi & =e^{1234}+e^{1256}+e^{1357}-e^{1467}-e^{2367}-e^{2457}+e^{3456}.
\end{align}
The full set of bilinears defined by $\epsilon_{\pm}$ is given in Table \ref{tab:Rank--gamma-matrix}.

\begin{table}
\begin{centering}
\begin{tabular}{ccccc}
\toprule 
$k$ & $\epsilon^{\transpose}_{+}\gamma_{(k)}\epsilon_{+}$ & $\epsilon^{\transpose}_{-}\gamma_{(k)}\epsilon_{-}$ & $\epsilon^{\transpose}_{+}\gamma_{(k)}\epsilon_{-}$ & $\epsilon^{\transpose}_{-}\gamma_{(k)}\epsilon_{+}$\tabularnewline
\midrule
\midrule 
0 & 1 & 1 & $\sin\zeta$ & $\sin\zeta$\tabularnewline
1 & $\bar{K}$ & $-\bar{K}$ & 0 & 0\tabularnewline
2 & 0 & 0 & 0 & 0\tabularnewline
3 & 0 & 0 & $-\bar{\phi}$ & $\bar{\phi}$\tabularnewline
4 & $Y$ & $Y$ & $\star Y$ & $\star Y$\tabularnewline
5 & $-\star\bar{\phi}$ & $\star\bar{\phi}$ & 0 & 0\tabularnewline
6 & 0 & 0 & 0 & 0\tabularnewline
7 & 0 & 0 & $-\star\bar{K}$ & $\star\bar{K}$\tabularnewline
8 & $\sin\zeta\vol_{8}$ & $\sin\zeta\vol_{8}$ & $\vol_{8}$ & $\vol_{8}$\tabularnewline
\bottomrule
\end{tabular}
\par\end{centering}
\caption{Rank-$k$ gamma matrix bilinears for the nowhere-vanishing Majorana spinors $\epsilon_{\pm}$ that define the $\protect\Gx 2$ structure on the eight-manifold $X$.\label{tab:Rank--gamma-matrix}}
\end{table}

\subsection{Torsion classes for a \texorpdfstring{$\Gx 2$}{G\_2} structure in eight dimensions\label{subsec:Eight-dimensions}}

In this appendix, we determine the form of the torsion classes of a $\Gx 2$ structure in eight dimensions by decomposing the exterior derivatives of the defining forms into irreducible $\Gx 2$ representations. 

Consider a $\Gx 2$ structure in eight dimensions. The intrinsic torsion takes values in~\cite{Martelli:2003ki}
\begin{equation}
2\ext^{0}_{\rep 1}\oplus4\ext^{1}_{\rep 7}\oplus2\ext^{2}_{\rep{14}}\oplus2\ext^{3}_{\rep{27}},\label{eq:int_torsion_space}
\end{equation}
where $\ext^{p}_{\rep r}$ indicates the space of $p$-forms transforming in the $\Gx 2$ representation $\rep r$, giving a total of ten modules. The corresponding torsion classes are determined by the exterior derivatives of $K$, $\phi$ and $\psi=\imath_{K}\star\phi$. A general decomposition of the exterior derivatives of a one-form, a three-form and a four-form gives
\begin{align}
\dd K & =K\wedge\tilde{\beta}_{\rep 7}+\gamma_{\rep 7}+\gamma_{\rep{14}},\\
\dd\phi & =K\wedge(\tilde{\delta}_{\rep 1}+\tilde{\delta}_{\rep 7}+\tilde{\delta}_{\rep{27}})+\alpha_{\rep 1}\psi+3\beta_{\rep 7}\wedge\phi+\star_{7}\delta_{\rep{27}},\\
\dd\psi & =K\wedge(\tilde{\tau}_{\rep 1}+\tilde{\tau}_{\rep 7}+\tilde{\tau}_{\rep{27}})+4\beta_{\rep 7}\wedge\psi+\eta_{\rep{14}},
\end{align}
where $\alpha_{\rep r}$ denotes a zero-form, $\beta_{\rep r}$ a one-form, $\gamma_{\rep r}$ a two-form, $\delta_{\rep r}$ a three-form, $\tau_{\rep r}$ a four-form, and $\eta_{\rep r}$ a five-form, while a tilde indicates a distinct form, so that, for example, $\tilde{\beta}_{\rep 7}\neq\beta_{\rep 7}$. The $\Gx 2$ representation structure implies that these forms satisfy a number of identities~\cite{math/0301218,math/0305124}:
\begin{align}
\begin{split}\gamma_{\rep 7}\wedge\phi & =2\star_{7}\gamma_{\rep 7},\\
\gamma_{\rep 7}\wedge\psi & =\star_{7}(\gamma_{\rep 7}\lrcorner\phi),\\
\gamma_{\rep{14}}\wedge\phi & =-\star_{7}\gamma_{\rep{14}},\\
\gamma_{\rep{14}}\wedge\psi & =0,\\
\delta_{\rep 1}\wedge\phi & =0,\\
\delta_{\rep 7}\wedge\psi\equiv\star_{7}\delta_{\rep 7}\wedge\phi & =0.
\end{split}
 & \begin{split}\delta_{\rep{27}}\wedge\phi=\delta_{\rep{27}}\wedge\psi\equiv\star_{7}\delta_{\rep{27}}\wedge\phi & =0,\\
\tau_{\rep 7}\wedge\phi\equiv\star_{7}\tau_{\rep 7}\wedge\psi & =0,\\
\star_{7}\tau_{\rep{27}}\wedge\phi=\star_{7}\tau_{\rep{27}}\wedge\psi\equiv\tau_{\rep{27}}\wedge\phi & =0,\\
\star_{7}\eta_{\rep 7}\wedge\phi & =2\eta_{\rep 7},\\
\star_{7}\eta_{\rep{14}}\wedge\phi & =-\eta_{\rep{14}},\\
\\\end{split}
\end{align}
Here the Hodge star on the seven-dimensional space transverse to $K$ is $\star_{7}=\imath_{K}\star$, so that $\psi=\star_{7}\phi$. These identities apply to both the tilded and untilded forms above.

Defining $\tilde{\delta}_{\rep 1}=3\tilde{\alpha}_{\rep 1}\phi$, a simple scaling argument then fixes $\tilde{\tau}_{\rep 1}=4\tilde{\alpha}_{\rep 1}\psi$. We also have the following $\Gx 2$-structure identity for the variation of $\psi=\star_{7}\phi$:
\begin{equation}
\delta\star_{7}\phi=\tfrac{4}{3}\star_{7}\delta_{\rep 1}\phi+\star_{7}\delta_{\rep 7}\phi-\star_{7}\delta_{\rep{27}}\phi.
\end{equation}
Together with the decomposition $\dd=\dd_{K}+\dd_{7}=K\wedge\partial_{K}+\dd_{7}$, this implies that the forms appearing in $\imath_{K}\dd\phi$ and $\imath_{K}\dd\psi$ must be related. Putting these together, one finds
\begin{align}
\dd K & =K\wedge\tilde{\beta}_{\rep 7}+\gamma_{\rep 7}+\gamma_{\rep{14}},\label{eq:torsion1}\\
\dd\phi & =K\wedge(3\tilde{\alpha}_{\rep 1}\phi+\tilde{\delta}_{\rep 7}+\tilde{\delta}_{\rep{27}})+\alpha_{\rep 1}\psi+3\beta_{\rep 7}\wedge\phi+\star_{7}\delta_{\rep{27}},\label{eq:torsion2}\\
\dd\psi & =K\wedge(4\tilde{\alpha}_{\rep 1}\psi+\star_{7}\tilde{\delta}_{\rep 7}-\star_{7}\tilde{\delta}_{\rep{27}})+4\beta_{\rep 7}\wedge\psi+\eta_{\rep{14}}.\label{eq:torsion3}
\end{align}
The independent torsion classes then span the space \eqref{eq:int_torsion_space}.

\section{Spinor bilinear conditions\label{app:bilinears}}

In this appendix, we collect the differential and algebraic identities satisfied by the $\Gx 2$-invariant spinor bilinears as a consequence of the Killing spinor equations. The Mathematica package \texttt{GammaMaP}~\cite{1905.00429} is very useful for computing the various differential and algebraic identities implied by the Killing spinor equations \eqref{eq:KSE-1}--\eqref{eq:KSE-2}. In particular, the differential conditions on the invariant forms defined by the $\Gx 2$ structure are
\begin{align}
\ee^{-3\Delta}\dd(\ee^{3\Delta}\sin\zeta) & =f-4m\bar{K},\nonumber \\
\dd(\ee^{3\Delta}\bar{K}) & =0,\nonumber \\
\ee^{-6\Delta}\dd(\ee^{6\Delta}\bar{\phi}) & =-\star F+F\sin\zeta-4mY,\nonumber \\
\ee^{-6\Delta}\dd(\ee^{6\Delta}Y) & =-F\wedge\bar{K},\label{eq:dY}\\
\ee^{-9\Delta}\dd(\ee^{9\Delta}\star Y) & =8m\star\bar{\phi}+f\wedge Y-\bar{K}\wedge\star F,\nonumber \\
\dd(\ee^{9\Delta}\star\bar{\phi}) & =0,\nonumber \\
\ee^{-12\Delta}\dd(\ee^{12\Delta}\star\bar{K}) & =-8m\sin\zeta\vol_{8},\nonumber 
\end{align}
where the invariant forms satisfy the identities
\begin{equation}
\begin{split}\begin{gathered}Y=-\psi+\sin\zeta\,\phi\wedge K,\qquad\star Y=-\sin\zeta\,\psi+\phi\wedge K,\\
\vol_{7}=-\star K=\tfrac{1}{7}\phi\wedge\psi,\qquad\vol_{8}=-\vol_{7}\wedge K.
\end{gathered}
\end{split}
\end{equation}
In addition, the external component of the gravitino equation yields further algebraic conditions. By taking suitable commutators and anticommutators, one finds 
\begin{equation}
\begin{split}0 & =2m-\tfrac{1}{3}\bar{K}\lrcorner f+\tfrac{1}{6}\star Y\lrcorner F,\\
0 & =2m\sin\zeta-\bar{K}\lrcorner\dd\Delta+\tfrac{1}{6}Y\lrcorner F,\\
0 & =2m\bar{K}+\sin\zeta\,\dd\Delta-\tfrac{1}{3}f+\tfrac{1}{6}\bar{\phi}\lrcorner F,\\
0 & =\dd\Delta-\tfrac{1}{3}f\sin\zeta-\tfrac{1}{6}\bar{\phi}\lrcorner\star F,\\
0 & =\dd\Delta\lrcorner\bar{\phi}+\tfrac{1}{3}\bar{K}\wedge f+\tfrac{1}{36}\star Y_{a}{}^{bcd}F_{bcde}e^{ae},\\
0 & =\bar{K}\wedge\dd\Delta+\tfrac{1}{3}f\lrcorner\bar{\phi}+\tfrac{1}{36}Y_{a}{}^{bcd}F_{bcde}e^{ae},\\
0 & =-\dd\Delta\lrcorner\star Y+\tfrac{1}{3}f\lrcorner Y+\tfrac{1}{24}\bar{\phi}_{a}{}^{bc}F_{bcde}e^{ade}+\tfrac{1}{6}\bar{K}\lrcorner\star F,\\
0 & =2m\bar{\phi}-\dd\Delta\lrcorner Y+\tfrac{1}{3}f\lrcorner\star Y-\tfrac{1}{6}\bar{K}\lrcorner F-\tfrac{1}{72}\star\bar{\phi}_{ab}{}^{cde}F_{cdef}e^{abf},\\
0 & =2m\star Y+\dd\Delta\lrcorner\star\bar{\phi}+\tfrac{1}{3}f\wedge\bar{\phi}+\tfrac{1}{6}F-\tfrac{1}{48}Y_{ab}{}^{cd}F_{cdef}e^{abef}+\tfrac{1}{6}\sin\zeta\star F,\\
0 & =2mY+\dd\Delta\wedge\bar{\phi}+\tfrac{1}{3}f\lrcorner\star\bar{\phi}+\tfrac{1}{6}\star F-\tfrac{1}{48}\star Y_{ab}{}^{cd}F_{cdef}e^{abef}+\tfrac{1}{6}\sin\zeta\,F,\\
0 & =2m\star\bar{K}+\sin\zeta\star\dd\Delta-\tfrac{1}{3}\star f+\tfrac{1}{6}\bar{\phi}\wedge\star F,\\
0 & =\star\dd\Delta-\tfrac{1}{3}\star f\sin\zeta-\tfrac{1}{6}\bar{\phi}\wedge F,\\
0 & =2m\vol_{8}-\tfrac{1}{3}\star\bar{K}\wedge f+\tfrac{1}{6}Y\wedge F,\\
0 & =2m\sin\zeta\vol_{8}-\star\bar{K}\wedge\dd\Delta+\tfrac{1}{6}\star Y\wedge F.
\end{split}
\label{eq:ext_8_1}
\end{equation}
The necessary and sufficient conditions derived in \cite{Martelli:2003ki}, and quoted in \eqref{eq:sufficient_1}--\eqref{eq:sufficient_6}, imply all of the differential and algebraic conditions listed above.

\subsection{Evaluating \texorpdfstring{$\epsilon_{-}^{\protect\transpose}\gamma^{m}\nabla_{m}\epsilon_{+}$}{the spinor derivative}\label{sec:Computing-the-superpotential}}

When evaluating the superpotential in terms of differential forms in \eqref{eq:Fierz_differential}, we needed to rewrite the term $\epsilon^{\transpose}_{-}\gamma^{m}\nabla_{m}\epsilon_{+}$ using a Fierz identity. We now give more details of this calculation.

Since $1=\epsilon^{\transpose}_{+}\epsilon_{+}$ and $\sin\zeta=\epsilon^{\transpose}_{+}\epsilon_{-}$ are scalar bilinears, we may insert either of them to rewrite $\epsilon^{\transpose}_{-}\gamma^{m}\nabla_{m}\epsilon_{+}$ in terms of the outer product $\epsilon_{+}\epsilon^{\transpose}_{+}$. For example, inserting ``$1$'' gives
\begin{align}
2^{4}\epsilon^{\transpose}_{-}\gamma^{m}\nabla_{m}\epsilon_{+} & =16(\epsilon^{\transpose}_{-}\gamma^{m}\nabla_{m}\epsilon_{+})(\epsilon^{\transpose}_{+}\epsilon_{+})\nonumber \\
 & =16\epsilon^{\transpose}_{-}\gamma^{m}\nabla_{m}(\epsilon_{+}\epsilon^{\transpose}_{+})\epsilon_{+}-16\cancel{\epsilon^{\transpose}_{-}\gamma^{m}\epsilon_{+}}\,\nabla_{m}\epsilon^{\transpose}_{+}\epsilon_{+}\nonumber \\
 & =16\epsilon^{\transpose}_{-}\gamma^{m}\nabla_{m}(\epsilon_{+}\epsilon^{\transpose}_{+})\epsilon_{+},\label{eq:deriv1}
\end{align}
where we have used the identity $\epsilon^{\transpose}_{-}\gamma^{m}\epsilon_{+}=0$ from Table \ref{tab:Rank--gamma-matrix}. If one instead multiplies by $\sin\zeta$, one finds
\begin{align}
2^{4}\epsilon^{\transpose}_{-}\gamma^{m}\nabla_{m}\epsilon_{+}\sin\zeta & =16(\epsilon^{\transpose}_{-}\gamma^{m}\nabla_{m}\epsilon_{+})(\epsilon^{\transpose}_{+}\epsilon_{-})\nonumber \\
 & =16\epsilon^{\transpose}_{-}\gamma^{m}\nabla_{m}(\epsilon_{+}\epsilon^{\transpose}_{+})\epsilon_{-}-16\cancel{\epsilon^{\transpose}_{-}\gamma^{m}\epsilon_{+}}\,\nabla_{m}\epsilon^{\transpose}_{+}\epsilon_{-}\nonumber \\
 & =16\epsilon^{\transpose}_{-}\gamma^{m}\nabla_{m}(\epsilon_{+}\epsilon^{\transpose}_{+})\epsilon_{-}.\label{eq:deriv2}
\end{align}
In both cases, this gives a relation between $\epsilon^{\transpose}_{-}\gamma^{m}\nabla_{m}\epsilon_{+}$ and a derivative of the outer product $\epsilon_{+}\epsilon^{\transpose}_{+}$. 

Using the Fierz identity \eqref{eq:fierz}, we can express $\epsilon_{+}\epsilon^{\transpose}_{+}$ in terms of the invariant forms and gamma matrices as
\begin{equation}
2^{4}\,\epsilon_{+}\epsilon^{\transpose}_{+}=\id_{16}+\gamma^{m}\bar{K}_{m}+\tfrac{1}{4!}\gamma^{mnpq}Y_{mnpq}-\tfrac{1}{5!}\gamma^{mnpqr}\star\bar{\phi}_{mnpqr}+\sin\zeta\,\gamma_{9}.
\end{equation}
Using this Fierz expansion in \eqref{eq:deriv1} gives
\begin{align}
16(\epsilon^{\transpose}_{-}\gamma^{m}\nabla_{m}\epsilon_{+}) & =\epsilon^{\transpose}_{-}\gamma^{s}\nabla_{s}(\cancel{\id_{16}}+\gamma^{m}\bar{K}_{m}+\tfrac{1}{4!}\gamma^{mnpq}Y_{mnpq}-\tfrac{1}{5!}\gamma^{mnpqr}\star\bar{\phi}_{mnpqr}+\sin\zeta\,\gamma_{9})\epsilon_{+}\nonumber \\
 & =\epsilon^{\transpose}_{-}\gamma^{s}\gamma^{m}\epsilon_{+}\nabla_{s}\bar{K}_{m}+\tfrac{1}{4!}\epsilon^{\transpose}_{-}\gamma^{s}\gamma^{mnpq}\epsilon_{+}\nabla_{s}Y_{mnpq}-\tfrac{1}{5!}\epsilon^{\transpose}_{-}\gamma^{s}\gamma^{mnpqr}\epsilon_{+}\nabla_{s}\star\bar{\phi}_{mnpqr}\nonumber \\
 & \eqspace+\nabla_{m}\sin\zeta\,\epsilon^{\transpose}_{-}\gamma^{m}\gamma_{9}\epsilon_{+}\nonumber \\
 & =\sin\zeta\,\nabla_{m}\bar{K}^{m}+\tfrac{1}{3!}\bar{\phi}_{npq}\nabla_{m}Y^{mnpq}-\tfrac{1}{4!}\star Y^{npqr}\nabla^{m}\star\bar{\phi}_{mnpqr}-\bar{K}^{m}\nabla_{m}\sin\zeta\nonumber \\
 & =\sin\zeta\,\nabla\lrcorner\bar{K}+\bar{\phi}\lrcorner(\nabla\lrcorner Y)-\star Y\lrcorner(\nabla\lrcorner\star\bar{\phi})-\bar{K}\lrcorner\dd\sin\zeta\nonumber \\
 & =\vol^{-1}_{8}(-\sin\zeta\,\dd\star\bar{K}-\bar{\phi}\wedge\dd\star Y-\star Y\wedge\dd\bar{\phi}+6\star\bar{K}\wedge\dd\sin\zeta)\nonumber \\
 & \equiv\vol^{-1}_{8}(-8\sin\zeta\,\dd\star\bar{K}-2\bar{\phi}\wedge\dd\star Y+6\star\bar{K}\wedge\dd\sin\zeta),\label{eq:spinor_deriv1}
\end{align}
in agreement with \eqref{eq:Fierz_differential}. Alternatively, substituting the Fierz identity into \eqref{eq:deriv2} gives
\begin{align}
16(\epsilon^{\transpose}_{-}\gamma^{m}\nabla_{m}\epsilon_{+})\sin\zeta & =\epsilon^{\transpose}_{-}\gamma^{m}\nabla_{m}(\cancel{\id_{16}}+\gamma^{m}\bar{K}_{m}+\tfrac{1}{4!}\gamma^{mnpq}Y_{mnpq}-\tfrac{1}{5!}\gamma^{mnpqr}\star\bar{\phi}_{mnpqr}+\sin\zeta\,\gamma_{9})\epsilon_{-}\nonumber \\
 & =\epsilon^{\transpose}_{-}\gamma^{s}\gamma^{m}\epsilon_{-}\nabla_{s}\bar{K}_{m}+\tfrac{1}{4!}\epsilon^{\transpose}_{-}\gamma^{s}\gamma^{mnpq}\epsilon_{-}\nabla_{s}Y_{mnpq}\nonumber \\
 & \eqspace-\tfrac{1}{5!}\epsilon^{\transpose}_{-}\gamma^{s}\gamma^{mnpqr}\epsilon_{-}\nabla_{s}\star\bar{\phi}_{mnpqr}+\nabla_{m}\sin\zeta\,\cancel{\epsilon^{\transpose}_{-}\gamma^{m}\epsilon_{+}}\nonumber \\
 & =\nabla^{m}\bar{K}_{m}+\tfrac{1}{4!}\star\bar{\phi}^{smnpq}\nabla_{s}Y_{mnpq}-\tfrac{1}{4!}Y^{npqr}\nabla^{s}\star\bar{\phi}_{snpqr}\nonumber \\
 & =\nabla^{m}\bar{K}_{m}+\star\bar{\phi}\lrcorner\dd Y-Y\lrcorner(\nabla\lrcorner\star\bar{\phi})\nonumber \\
 & =\vol^{-1}_{8}(-\dd\star\bar{K}-\bar{\phi}\wedge\dd Y-Y\wedge\dd\bar{\phi})\nonumber \\
 & \equiv\vol^{-1}_{8}(6\dd\star\bar{K}-2Y\wedge\dd\bar{\phi}).\label{eq:spinor_deriv2}
\end{align}
Although \eqref{eq:spinor_deriv1} and \eqref{eq:spinor_deriv2} appear at first sight to give different expressions, their agreement reduces to expressing them in terms of $K$, $\phi$ and $\psi$, and using the identity
\begin{equation}
\phi\wedge\dd\psi=-\tfrac{4}{3}\dd\phi\wedge\psi,
\end{equation}
which is implied by the expressions for $\dd\phi$ and $\dd\psi$ in \eqref{eq:torsion2} and \eqref{eq:torsion3}. Either way, one finds
\begin{equation}
16\vol_{8}(\epsilon^{\transpose}_{-}\gamma^{m}\nabla_{m}\epsilon_{+})=-\tfrac{8}{7}\dd\zeta\wedge\phi\wedge\psi-2\cos\zeta\,\phi\wedge K\wedge\dd\phi.
\end{equation}

If instead we had multiplied $\epsilon^{\transpose}_{-}\gamma^{m}\nabla_{m}\epsilon_{+}$ by $1=\epsilon^{\transpose}_{-}\epsilon_{-}$ or $\sin\zeta=\epsilon^{\transpose}_{-}\epsilon_{+}$, we would have needed the Fierz identity
\begin{equation}
2^{4}\,\epsilon_{+}\epsilon^{\transpose}_{-}=\sin\zeta\id_{16}-\tfrac{1}{3!}\gamma^{mnp}\bar{\phi}_{mnp}+\tfrac{1}{4!}\gamma^{mnpq}\star Y_{mnpq}-\tfrac{1}{7!}\gamma^{mnpqrst}\star\bar{K}_{mnpqrst}+\gamma_{9}.
\end{equation}
Carrying out the same manipulations as above, one again finds \eqref{eq:spinor_deriv1} or \eqref{eq:spinor_deriv2}.

\bibliographystyle{utphys}
\bibliography{inspire, extra}

\end{document}